\documentclass[acmsmall]{acmart}

\usepackage{listings}
\usepackage[linesnumbered ]{algorithm2e}
\usepackage{amsmath}
\usepackage{xcolor}
\usepackage{tikz}
\usetikzlibrary{shapes,arrows,quotes,positioning,calc,fit,backgrounds}

\usepackage{multirow}

\AtBeginDocument{%
  \providecommand\BibTeX{{%
    \normalfont B\kern-0.5em{\scshape i\kern-0.25em b}\kern-0.8em\TeX}}}

\setcopyright{acmlicensed}
\acmJournal{TOSEM}
\acmYear{2020} \acmVolume{1} \acmNumber{1} \acmArticle{1} \acmMonth{1} \acmPrice{15.00}\acmDOI{10.1145/3424307}

\begin{document}


\title{Adversarial Specification Mining}

\author{Hong Jin Kang}

\email{hjkang.2018@phdis.smu.edu.sg}
\affiliation{%
  \institution{School of Information Systems, Singapore Management University}
  \streetaddress{80 Stamford Rd}
  \city{Singapore}
  \postcode{178902}
}

\author{David Lo}
\affiliation{%
\institution{School of Information Systems, Singapore Management University}
\streetaddress{80 Stamford Rd}
\city{Singapore}
\postcode{178902}}
\email{davidlo@smu.edu.sg}


\begin{abstract}
  There have been numerous studies on mining temporal specifications from execution traces. 
  These approaches learn finite-state automata (FSA) from execution traces when running tests. 
  To learn accurate specifications of a software system, many tests are required. 
  Existing approaches generalize from a limited number of traces or use simple test generation strategies.
  Unfortunately, these strategies may not exercise uncommon usage patterns of a software system.
  To address this problem, we propose a new approach, 
  adversarial specification mining, and develop a prototype, DICE (Diversity through Counter-Examples). 
  DICE has two components: DICE-Tester and DICE-Miner.
  After mining Linear Temporal Logic specifications from an input test suite,
  DICE-Tester adversarially guides test generation, searching for counterexamples to these specifications to invalidate spurious properties.
  These counterexamples represent gaps in the diversity of the input test suite.
  This process produces execution traces of usage patterns that were unrepresented in the input test suite.
  Next, we propose a new specification inference algorithm, DICE-Miner, to infer FSAs using the traces, guided by the temporal specifications. 
  We find that the inferred specifications are of higher quality than those produced by existing state-of-the-art specification miners. 
  Finally, we use the FSAs in a fuzzer for servers of stateful protocols, increasing its coverage.

\end{abstract}




\maketitle

\section{Introduction}

In an ideal world, software systems and their APIs should be clearly documented to prevent misuse and bugs. 
In practice, software systems are usually released without documentation and existing documentation may become out of date as the software system evolves~\cite{zhong2013detecting}. 
The lack of specifications causes difficulties in program comprehension,  
and the misuse of APIs has been recognized as a leading cause of software bugs and vulnerabilities~\cite{nadi2016jumping,egele2013empirical}. 

To address the lack of specifications, researchers have proposed many techniques to automatically infer specifications and usage models, frequently in the form of Finite State Automata (FSA). 
These techniques require execution traces of the software as input. 
It is assumed that these traces are representative of the software and that all correct behavior are reflected in these traces.
Unfortunately, it was found that automatically mined specifications are still inaccurate~\cite{legunsen2016good}.
One reason for this may be that the traces used to construct these specifications are not representative and are not sufficiently diverse. 
Other researchers have proposed techniques~\cite{cohen2015have,busany2016behavioral} to determine if enough traces have been seen, 
but these techniques only consider the traces that have already been seen 
and metrics are computed only over the observed traces.
They are unable to reason about execution traces which are possible but are uncommon. 

Test generation may be a way to generate new tests that specification miners can learn from, 
and in several studies, researchers have used test generation to mine specifications. 
Tautoko~\cite{dallmeier2010generating}, for example, refines a FSA-based specification by generating tests to cover missing transitions.
Deep Specification Miner (DSM)~\cite{le2018deep} leverages random test generation to produce a large number of traces to learn language models from.
Still, as we investigate later in this study, these techniques are not sufficient for producing highly accurate models.
DSM relies on randomized test generation, and even when provided with traces of uncommon usage, it is not able to leverage these traces to produce more accurate models. 
Tautoko relies on methods that reveal the state of an object to detect a state change, 
which may limit it from working effectively for all types of objects.
As such, we hypothesize that existing test generation strategies do not completely address the problem of uncommon usage patterns. 

For ensuring that uncommon usage is represented, 
we propose a process which we term \textit{adversarial specification mining}.
In the first phase, we mine specifications from traces collected from running a set of test cases of the software under test.
In the second phase, test generation is guided towards the discovery of counterexamples of the mined specifications.
In the third phase, a specification miner uses the new counterexamples to construct an accurate model.
We developed a prototype, \textit{DICE} (\textbf{Di}versity through \textbf{C}ounter-\textbf{E}xamples). 
DICE mines FSA models through an adversarial specification mining process.
For the purpose of inferring a more accurate model, DICE produces more example execution traces 
given an initial set of temporal specifications, aiming to find inaccuracies in them. 
This is done through a search-based test generation process is adversarial to the input specifications, 
searching for tests that exercise the software under test in ways that the input specifications would not accept as correct usage. 
DICE contains two main components: \textbf{DICE-Tester}, which drives test generation towards uncommon patterns, 
and \textbf{DICE-Miner}, which converts execution traces into a Finite-State Automata (FSA) model.
This is the first study that makes use of search-based testing for mining specifications. 

DICE-Tester uses a search-based testing framework, Evosuite\cite{fraser2011evosuite}, 
guiding it towards the generation of counterexamples of the input specifications
by representing traces that will falsify the specifications as search goals.
The modifications made by DICE-Tester prevents the search algorithm in Evosuite from getting caught in a local optima, 
enabling Evosuite to efficiently search for counterexamples. 
We use the DynaMOSA algorithm, introduced in a previous study~\cite{panichella2017automated}, to allow Evosuite to dynamically 
select objectives instead of trying to achieve pareto optimality. 

To characterize a set of traces, 
we first mine specifications as properties in Linear Temporal Logic (LTL), a formalism of constraints on event-ordering, 
that hold on the traces. 
Prior work has shown the relationship between LTL and specification mining~\cite{dwyer1999patterns,le2015synergizing,beschastnikh2014using,beschastnikh2011leveraging,yang2006perracotta},
and has applied data mining techniques to infer temporal properties~\cite{le2015beyond,cao2018rule}.
Still, automatically mined properties are typically not completely accurate.
As such, this motivates work on boosting the accuracy of identifying temporal properties. 
Adversarial specification mining solves this problem by filtering out temporal properties that can be invalidated.
In this study, we use six LTL property templates introduced in previous studies~\cite{le2015synergizing,sun2019mining,beschastnikh2014using}.
However, we propose a reformulation of three properties, in which we use knowledge of method purity derived from low-cost heuristics and static analysis, 
to reformulate the properties, tackling shortcomings of the properties described in a recent study~\cite{sun2019mining}. 
In this work, we use the terms ``pure'' and ``side-effect-free'' interchangeably. 

As a result of guiding test generation to search for counterexamples, 
the traces collected by DICE-Tester include uncommon, but correct, usage patterns of the library. 
To infer an FSA model, we use the traces and temporal properties in a FSA inference algorithm that we propose. 
We borrow insights from prior work~\cite{de2010automated,krka2014automatic}, 
characterizing the states in an FSA based on the methods that are enabled and which can be invoked from it. 
We make two observations of limitations in existing model inference algorithms and modify our algorithm to address them.
In our evaluation of DICE, we find that the models produced by DICE outperform models from existing specification miners, 
such as the state-of-the-art specification miner, Deep Specification Miner~\cite{le2018deep}. 
Finally, we compare DICE and DSM by using the models they infer in server fuzzing, and we find that the models learned by DICE
helps in increasing line and branch coverage on an FTP server.

This paper makes the following contributions:
\begin{itemize}
  \item We propose the approach of adversarial specification mining. 
  Our prototype of this approach, DICE, produces a Finite-State-Automata from an initial test suite.
  This process involves two main components, a test generation process identifying counterexamples, and a state-of-the-art FSA specification mining process.
 \item We are the first to propose the use of an adversarial, search-based test generation process to produce diverse examples to learn from. 
 \item We propose a new specification miner that uses LTL properties, and avoids weaknesses of existing approaches, allowing for the inference of higher-quality Finite State Automata. 
 \item We empirically evaluate DICE against multiple baseline approaches, including state-of-the-art tools.
 \item We use the state machine output from DICE  in a stateful server fuzzer proposed in prior work, and we show that DICE and the learned state machines can help improve fuzzing.
\end{itemize}

The rest of the paper is organized as follows. 
Section \ref{sec:background} introduces the background of this study.
Section \ref{sec:approach} presents DICE in detail.
Section \ref{sec:evaluation} empirically compares DICE against other specification mining tools, answering several research questions of this study.
Section \ref{sec:discussion} provides some discussion of our work, including a qualitative evaluation, its practical applicability in server fuzzing, and threats to validity.
Section \ref{sec:related_work} discusses studies related to our work.
Section \ref{sec:conclusion} presents our conclusions and future directions of our work 

\section{Background}
\label{sec:background}

\subsection{Specification Mining}

\subsubsection{K-tails and its variants}

Many specification mining algorithms have been proposed.
To infer FSA, specification mining algorithms have to provide abstractions over states, 
and determine if two traces result in the same state.  
A classic algorithm that infers a Finite State Automaton from traces is the k-tails algorithm~\cite{ammons2002mining}.
The k-tails algorithm~\cite{ammons2002mining} first uses the input execution traces to build a Prefix Tree Acceptor (PTA). 
A PTA is a tree-like deterministic finite automaton (DFA) where states are grouped and merged based on the prefix that they share. 
This automaton is consistent with the input traces and will accept all of them. 
Next, the algorithm merges states that have the same sequences of invocations 
in the next k steps.
The value of the parameter, k, can vary. 
This trades off precision and recall; a small value of k results in more spurious merges while a large value of k leads to lower generality.

Studies have extended the traditional k-tails algorithm. 
These studies often keep the first step of the original k-tails algorithm,  
using a PTA to create a tree-like automaton that directly represents the input traces.
These algorithms thus inherit the assumptions made by k-tails in its first step, 
that states with the same prefix are equivalent, 
typically only modifying the second step of the k-tails algorithm, 
changing the equivalence criteria of states before merging them.

Lo et al.~\cite{lo2009automatic} propose to mine temporal rules that hold over the input traces and prevent any merge that will result in a violation of the rules.
Lorenzoli et al.~\cite{lorenzoli2008automatic} introduce GK-tail, which mines extended FSA where transitions are labelled not only with method calls, but includes parameter values. 
They introduce different merging criteria, 
including criteria that do not require exact matches of the transitions, and allow for more general conditions of the parameter values.
Krka et al.~\cite{krka2014automatic} introduce multiple algorithms in their work, including SEKT, which extends k-tails by adding another condition for equivalence: 
States are merged only if they correspond to the same abstract state, which are defined by the invariants extracted by Daikon.
Le et al.~\cite{le2018deep} propose a deep-learning based approach, Deep Specification Miner (DSM), to determine if a set of states are equivalent. 
They train a Recurrent Neural Network-based model to produce features characterizing each state 
in a high-dimensional space. 
After clustering the states in this space, states are merged according to the clusters they belong to. 
Therefore, each cluster is mapped to a single state in the output FSA. 
In this study, states are characterized by a feature vector built for each state, 
which includes the likelihood of each possible transition label based on their prefix.

\subsubsection{State abstraction}
\label{sec:state_abstraction}

While k-tails and its variants combine states based on their prefixes and an equivalence criteria,  
other approaches have proposed other methods to infer the states in an FSA.
de Caso et al.~\cite{de2010automated} propose CONTRACTOR, which uses program invariants to characterize states of an FSA based on the \textit{enabledness} of methods. 
A method is enabled if the invariants of the state hold. 
States are thus a combination of enabled methods, where the pre-conditions of the methods are consistent with one another.
CONTRACTOR was proposed as a method to validate pre- and post-conditions specifications by presenting a state machine abstraction 
of the specifications. 
The finite state machine help in revealing potential inaccuracies among the pre- and post-conditions.
Constructing all possible combinations of enabledness of the methods result in number of states exponential to the number of methods.
To avoid this state space blowup, CONTRACTOR models the dependencies between method enabledness 
to reduce the number of states. 
Afterwards, only the states reachable from the initial state are retained.
Krka et al.~\cite{krka2014automatic} enhance the CONTRACTOR model by proposing CONTRACTOR++, 
filtering invariants inferred by Daikon~\cite{ernst2007daikon} and including the output value of method invocations in the labels of transitions.

Finally, 
approaches such as ADABU~\cite{dallmeier2006mining} and Tautoko~\cite{dallmeier2010generating} identify a set of inspectors for each class. 
Inspector methods are heuristically identified based on their return type (not void), a lack of parameters, and the lack of side-effects.
Abstract states are characterized by the return values of these inspector methods, which are abstracted over to prevent a large number of states.
For example, an integer return value is abstracted into one of three abstract values based on its relative value to 0 (either $>$0, $=$ 0, $<$ 0).
These approaches may not perform well in the absence of inspectors.

\subsubsection{Temporal properties}

Several techniques have shown the use of temporal properties in inferring FSA from traces~\cite{dwyer1999patterns,le2015synergizing,beschastnikh2014using,beschastnikh2011leveraging,yang2006perracotta}.
As mentioned earlier, Lo et al.~\cite{lo2009automatic} use temporal properties to prevent erroneous merges. 
Data mining has been used for the identification of these rules; however, the number of false positives of inferred rules can be high, 
motivating the need for better ways to identify temporal rules~\cite{thummalapenta2009mining}.
Le et al.~\cite{le2015beyond} have studied the use of different interestingness measures, 
while Cao et al.~\cite{cao2018rule} proposed the use of learning-to-rank algorithms composing different interestingness measures to identify accurate properties.
Le et al.~\cite{le2015synergizing} have also built a meta-model, SpecForge, over existing algorithms in order to decompose mined FSAs into temporal rules, and recompose selected ones back into an FSA. 
In recent work, Sun et al.~\cite{sun2019mining} used crowdsourcing for identifying correct temporal properties, however, this process 
was not done automatically, relying on human annotators.

\subsection{Test Generation for Specification Mining}
Test generation for specification mining have been studied previously. 
Xie and Notkin~\cite{xie2003mutually} propose a feedback loop between specification inference and test generation. 
However, there is no publicly available version of a tool that implements this strategy and this strategy was not empirically evaluated.  
Tautoko~\cite{dallmeier2010generating} uses test generation to further refine a specification. 
Tautoko mutates an initial test suite and a given FSA model to find missing transitions in the FSA model. 
DSM~\cite{le2018deep} uses Randoop~\cite{pacheco2007randoop}, which performs randomized test generation, and traces are collected from the test cases generated  
to train a Recurrent Neural Network on.

The above studies use randomized testing or mutate an existing test suite for mining specifications. 
These test generation techniques do not systematically diversify the test suite or 
use any strategy to ensure sufficient diversity in the test cases.

\subsection{Search-based test generation}

In this study, we use search-based test generation to create test cases to learn specifications from. 
The generation of test cases are guided towards search goals that we define.  
We opt to use a search-based test generation tool, Evosuite~\cite{fraser2011evosuite}. 
Evosuite is a unit test generation tool for Java that uses a evolutionary approach to search for high-quality test cases 
that fulfils a specified set of coverage criteria.
Evosuite evolves a population of tests through multiple generations, and in each generation,
discards tests that are less fit while mutating surviving tests. 
This acts as a search process that iteratively improve the test cases to cover the search objectives. 
Many optimizations have been proposed and implemented in Evosuite since its inception~\cite{arcuri2013parameter,rojas2016seeding}. 
Evosuite is automated and does not require any manually written tests as input.  
Developers can extend the search algorithm or add new coverage and fitness goals. 
Evosuite comes with a variety of coverage goals, ranging from structural coverage to method coverage goals.
Structural coverage goals include line coverage and branch coverage, while method coverage goals 
guide Evosuite towards tests that invoke every constructor and method of the class. 
We select Evosuite instead of alternative tools due to its strong performance among state-of-the-art test generation tools~\cite{molina2018java}.

\subsubsection{Multiple objective formulation} In the past, test case generation focused on optimising for various coverage criteria independently of each other.
Recently, Rojas et al.~\cite{rojas2015combining} generated tests while optimising multiple objectives simultaneously, 
aggregating fitness functions through a weighted sum. 
However, other studies show the limitations of aggregating multiple fitness goals as a single measure.
For example, one such limitation is that the weighted sum aggregation assumes that each fitness goal is independent of each other, 
which is not true of structural coverage goals (for example, conditional dependencies mean that line and branch coverage goals may depend on one another).
Instead of optimizing tests towards a single aggregated fitness value, 
other researchers have applied multi-objective search algorithms~\cite{panichella2015reformulating}.
These algorithms presents several advantages, including preventing the search process from getting stuck in a local minima, 
and can generate high quality test cases~\cite{panichella2015reformulating}.
Indeed, Gay~\cite{gay2017fitness} showed that optimising for multiple objectives at the same time instead of enumerating through
the objectives one by one lead to test suites that better detect faults.

\begin{figure}[t]
	\centering
	
	\includegraphics[width=0.2\linewidth]{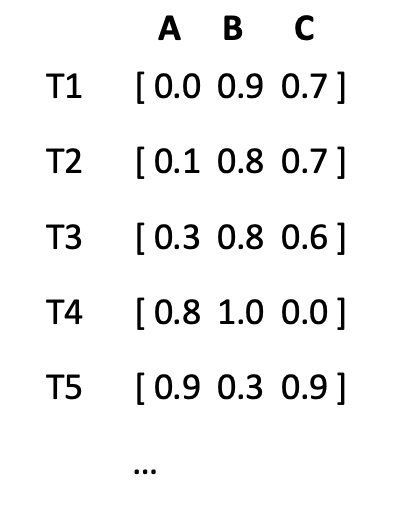}
	\caption{Example of several objective function vectors of test cases that are non-dominated. There may be more than a few non-dominated test cases and each of them have an equal chance to get included. We use DynaMOSA to address this problem.}
	\label{fig:dominated}
\end{figure}

There are problems specific to test generation when formulated as a multi-objective search.
When faced with a large number of search goals,
it is impossible to rank many of the individual test cases when considering all of the goals.
Due to this, the search process may degrade to become essentially random~\cite{panichella2015reformulating}. 
Another problem is that a test case that may be fit, 
when considering every search goal, 
but may not fully cover any individual search goal.
In other words, although the multi-objective formulation of test generation may produce test cases that are pareto-optimal, 
with the tests representing optimal trade-offs between fitness goals, 
it may not produce a resulting test suite that completely covers an objective.
This is detrimental to our study as we require test cases that contradict a temporal specification, instead of just 
\textit{being close} to covering it, 
regardless of the number of other fitness goals the tests are close to covering. 
Such a set of test cases will provide us with no value. 
For example, given several test cases which are scored as the vectors shown in Figure \ref{fig:dominated}, 
all the tests are non-dominated 
(each test is no worse than another with respect to at least one search goal).
In this example, the individual values in the vector represent the distance for a particular search goal.
A lower distance is better and a distance of 0.0 indicate that the test case covers that goal.
While objective A is covered by test T1 and objective C by T4, objective B is not covered by any test. 
As none of the tests dominate each other, they have equal probability of getting selected for the next generation.
Objective B is a difficult objective to cover. 
In our study, it is important that we retain and evolve test case T5 in the next generation as it is closest to covering objective B.

\begin{algorithm}
  \SetAlgoLined
  \KwIn{A set of coverage goals $C$.}
  \KwIn{Program, $P$.}
  \KwIn{Population Size $M$.}
  \KwOut{A test suite, $TS$, which is a collection of test cases}

  D = GetControlDependencies(P)

  P = RandomPopulation();

  A = InitArchive(P, C);

  C' = UpdateCurrentGoals(A, C, D)\;

   \While{Search budget is not expanded} {
     Q = GenerateOffspring(P)\;

     P = \{\}\; 

     A = UpdateArchive(A, Q, C')\;

     C' = UpdateCurrentGoals(A, C', D)\;

     fronts = Rank(Q)\;

     \For{front $\gets$ fronts} {
      \If{P.size >= M}{
          break\;
      }
      \For{$TC$ $\gets$ fronts} {
        \If{P.size + 1 > M}{
          break\;
        }
        AddToPopulation(P, $TC$);
      } 

     }
     
   }

   TS = A.getTestCases()\;
   
   \caption{Simplified version of the DynaMOSA algorithm. Given a program, $P$, and the set of coverage goals, $C$, DynaMOSA constructs a test suite, $TS$.}
   \label{algo:dynamosa}
\end{algorithm}

To address these problems, the DynaMOSA multi-objective algorithm~\cite{panichella2017automated} has been proposed for Evosuite.
The DynaMOSA algorithm, at a high-level, is given in Algorithm \ref{algo:dynamosa}.
It evolves an initial randomly generated population of tests through multiple generations. 
Given a set of coverage goals, the algorithm evolves a population of test cases through the usual mutation and cross-over operators (line 5). 
This gives us the offspring, $Q$, a set of test cases containing new tests as well as retaining some test cases from the previous generation.
The test cases in $Q$ are ranked and binned into a list of fronts, which partitions $Q$. 
The first front contains the best test case with respect to each coverage goal.
After the first front, each subsequent front ranks the remaining test cases by their pareto-optimality. 
The length (number of statements) of the test case is used as a tiebreaker when two test cases have the same score, 
preferring shorter test cases which is more likely to run in shorter time.
Then, the top ranked test cases form the population of the next generation and the offspring of this population are generated, 
and the process continues until the search budget is exhausted.

\subsubsection{Archive} DynaMOSA maintains an archive, $A$ (used in lines 8 and 23), similar to other search-based test generation strategies. 
During the test generation process, the archive stores test cases covering previously uncovered goals and provides a way to retrieve 
the best test case for a particular search goal.
The archive accounts for accidental coverage; a goal may be collaterally covered by a previous search for another set of goals.
When a test case for a particular search goal is stored in the archive, this search goal is removed from the current set of goals (line 8). 
As a consequence, the current set of search goals contains only the uncovered goals and focuses the search process on them.
The archive is updated whenever the search goal is covered, but also when a test case that is shorter than the current test 
and covers the same goal. 
At the end of the test generation process, the test cases in the archive are retrieved and are the output test suite (line 22).

\begin{figure}[h]
	\centering
	\scriptsize{
\begin{lstlisting}[language=java,numbers=left,escapechar=!,basicstyle=\ttfamily]
if (functionA())) {  // Initially targeted as it does not depend on another line
  if (functionB()) { // Targeted only after line 1 is covered
    functionC();     // Targeted only after lines 1 and 2 are covered
  }
} else {
  functionD();       // Targeted only after line 1 is covered
}
\end{lstlisting}
    \caption{DynaMOSA uses the control dependencies to dynamically target only search goals that can be covered. Line 3 is targeted only after lines 1 and 2 are covered. After lines 1 and 2 are covered, line 3 is added to the current set of goals.}
    \label{fig:dynamosa_control_dependencies}
	}
\end{figure}

\subsubsection{Dynamic selection of targets} The key feature of DynaMOSA is that it allows Evosuite to dynamically select targets 
based on the control dependencies between one another.
Initially, only a subset of the coverage goals that are independent of other goals are targeted.
Dynamically selecting targets allows Evosuite to be more efficient when trying to cover multiple structural goals.

For example, statements within branches require the if-statement to be covered first, 
therefore these statements are initially not targeted by Evosuite until the if-statement has been covered. 
If the if-statement has not been covered, then these goals cannot be covered.
The fitness, 
of a test with respect to these goals is, therefore, always worse than the fitness of the test with respect to goal of covering the if-statement. 
In the example shown in Figure \ref{fig:dynamosa_control_dependencies} statement 3 cannot be covered before both statements 1 and 2 are covered.
If a test case has not covered statement 1, it is not necessary to consider the fitness value of a test with respect to 
the search goal of covering statement 3.
Therefore, these uncoverable goals do not contribute meaningfully to the ranking. 

Evosuite first computes a control dependency graph of the program before generating tests and uses information 
from the control dependency graph to update the current set of search goals.
As described earlier,
before ranking test cases by their pareto-optimality, 
DynaMOSA first ensures that tests that are closest to a targeted search goal always survive and are retained in the next generation, 
even if these tests are not pareto-optimal with respect to the other goals.
This makes it more probable for Evosuite to progress towards individual goals, including those that may be difficult to cover. 
This particular feature is the reason why we build DICE on top of the DynaMOSA strategy. 

In Algorithm \ref{algo:dynamosa}, the dynamic selection of targets can be seen in lines 4 and 9, in which the archive is used to 
determine which goals have been covered. 
DynaMOSA uses the control dependencies, $D$, to determine the initial set of targets in line 4.
As the goals are covered, it adds the goals that depend on the covered goals in line 8 to the currently targeted set of goals, $C'$.

\section{DICE Approach}
\label{sec:approach}

\subsection{Overview}

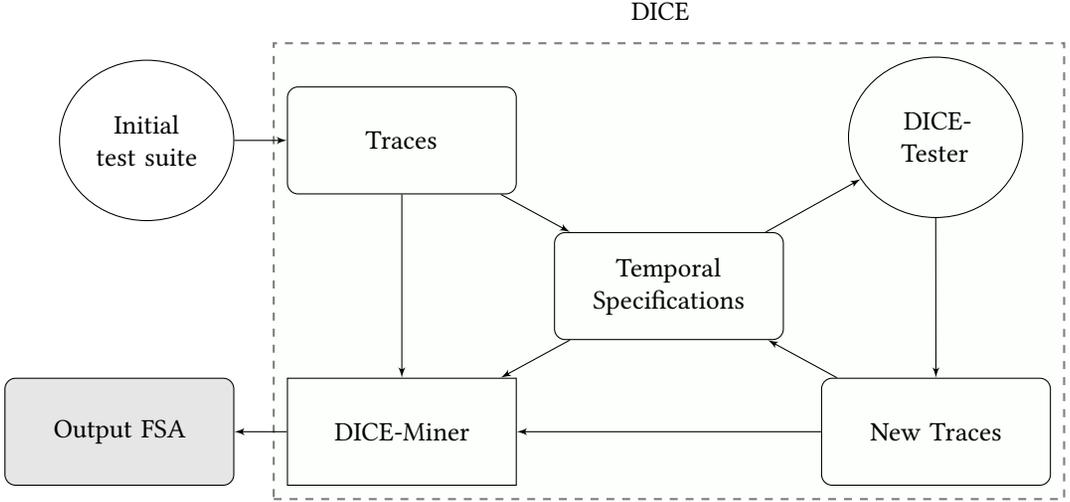
\begin{figure*}
  \begin{center}
    
  \begin{tikzpicture}[node distance = 0.7cm , yscale=0.5,
    ,line/.style={draw, -latex'},
    test/.style=%
      {%
        ellipse, draw, text width=4em, text centered, rounded
        corners, minimum height=6em, minimum width=4em
      },
      trace/.style=%
      {%
      rectangle, draw, text width=8em, text centered, rounded
        corners, minimum height=4em, minimum width=4em
      },
      fsa/.style=%
      {%
      rectangle, draw, fill=gray!20,  text width=8em, text centered, rounded
        corners, minimum height=4em, minimum width=4em
      },
      outer/.style={draw=gray,dashed,fill=green!1,thick,inner sep=5pt },
      drawtext/.style={opacity=0.0,text opacity=1, minimum height=4em},
      miner/.style=%
      {%
      rectangle, draw,   text width=8em, text centered, minimum height=4em, minimum width=4em
      }]
      
      \node [test] (init_test) {Initial test suite};
      \node [trace,right= of init_test] (init_traces) {Traces};
      \node [trace,below right= of init_traces] (temporal_specifications) {Temporal Specifications};
      \node [test, above right=of temporal_specifications, xshift=0.7cm] (testing) {DICE-Tester};
      \node [trace, below right =of temporal_specifications ] (new_traces) {New Traces};

      \node [miner, below left =of temporal_specifications] (spec_miner) {DICE-Miner};
      
      \node [fsa, left =of spec_miner] (output) {Output FSA};

      \path [line] (init_test) -- (init_traces) ;
      
      \path [line] (init_traces)  -- (temporal_specifications);
      \path [line] (temporal_specifications)  -- (spec_miner);
      \path [line] (temporal_specifications)  -- (testing);
      \path [line] (testing) -- (new_traces);
      \path [line] (new_traces) -- (spec_miner);
      \path [line] (new_traces) -- (temporal_specifications);
      \path [line] (init_traces) --(spec_miner);
      \path [line] (spec_miner) -- (output);
  
      \begin{pgfonlayer}{background}
        \node[outer,fit=(init_traces) (testing) (new_traces) (spec_miner)] (background_0) {};
      \end{pgfonlayer}

      \node[drawtext,text width=4cm, above= of background_0, xshift=1.5cm, yshift=-1cm]  {DICE};

  \end{tikzpicture}
    
  \end{center}
  \caption{High-level overview of DICE}
  \label{fig:approach}
\end{figure*}

We show a high-level overview of the approach used by DICE in Figure \ref{fig:approach}. 
DICE consists of 2 main components: DICE-Tester and DICE-Miner. 
From a high-level perspective, DICE takes a class under test and an initial test suite as input, producing a FSA model as output. 
DICE first exercises the test suite, collecting the execution traces. 
This is followed by three phases: 
\begin{itemize}
\item \textbf{Mining Purity-Aware Temporal Specification.} First, temporal specifications, in the form of LTL temporal properties, are mined from these traces while being aware of method purity. 
\item \textbf{Adversarial Test Generation.} The temporal specifications are fed into DICE-Tester and are converted into search goals for the test generation process.
DICE-Tester is adversarial to the temporal specifications and refines them by invalidating incorrect properties, while generating new test cases and collecting the execution traces of these test cases.
\item \textbf{FSA Inference.} Finally, the new traces and temporal specifications, with invalid specifications now removed, are input into DICE-Miner, which will infer an FSA.
      DICE-Miner avoids weaknesses of existing algorithms by using method purity and the temporal specifications to prevent over-generalisation.

\end{itemize}

\subsection{Mining Purity-Aware Temporal Specification}

The test generation phase of the adversarial specification mining process requires a set of specifications.
As such, the first phase of DICE is to mine temporal specifications over the input traces collected from the input test suite.  
A formalism over constraints ordering events is Linear Temporal Logic (LTL)~\cite{pnueli1977temporal,huth2004logic}. 
Like previous studies in specification mining~\cite{dwyer1999patterns,le2015synergizing,beschastnikh2014using,beschastnikh2011leveraging,yang2006perracotta}, we use LTL to specify constraints over events. 
In this work, events are specifically method invocations of a class and the following subset of LTL connectives are used in the property templates:
\begin{enumerate}
  \item X $\phi$ means that $\phi$ has to hold at the ne\textbf{X}t state.
  \item F $\phi$ means that $\phi$ has to hold at some \textbf{F}uture state.
  \item G $\phi$ means that $\phi$ has to hold \textbf{G}lobally at all future states.
  \item $\rho$ U $\phi$ is `\textbf{U}ntil', which means that $\phi$ has to hold at some point. $\rho$ has to hold until $\phi$ holds. 
  \item $\rho$ W $\phi$ is `\textbf{W}eak until', which means that $\rho$ has to hold until $\phi$ holds. If $\phi$ never becomes true, $\rho$ has to hold forever.
\end{enumerate}

Six LTL property templates are commonly used in previous studies. 
The six LTL property templates are described as follows:
\begin{enumerate}
  \item AF(a, b): an occurrence of event a must be eventually followed by event b. In LTL, this rule is $G(a \rightarrow XF b)$
  \item NF(a, b): an occurrence of event a is never followed by event b. In LTL, this rule is $G(a \rightarrow XG(\neg b))$
  \item AP(a, b): an occurrence of event a must be preceded by event b. In LTL, this rule is $\neg a W b$
  \item AIF(a, b): an occurrence of event a be immediately followed event b. In LTL, this rule is $G(a \rightarrow b)$
  \item NIF(a, b): an occurrence of event a is never immediately followed by event b. In LTL, this rule is $G(a \rightarrow X(\neg b))$
  \item AIP(a, b): an occurrence of event a must be immediately preceded by event b. In LTL, this rule is $F(a) \rightarrow (\neg a~U (b \land  X a))$
\end{enumerate}

The last three properties, introduced by Beschastnikh et al.~\cite{beschastnikh2014using} 
and Le et al.~\cite{le2015synergizing}, are "immediately" variants of the first three properties and 
have been shown to be useful for describing FSAs.
In this work, we use the LTL property templates from previous work which only consider 2 events. 
Later, in Section 5.4 (Qualitative Evaluation), we note some limitations of considering only 2 events at a time. 
We also note that the primary objective of this study is to infer automata models, 
and the LTL specifications are only later used to guide the testing process and the inference of the models.

While we use the same LTL property templates studied by Beschastnikh et al.~\cite{beschastnikh2014using}  and Le et al.~\cite{le2015synergizing}, 
in our work, we adapt three of them.
Recently, Sun et al. \cite{sun2019mining} pointed out shortcomings of these patterns when using crowdsourcing to identify temporal specifications. 
AIP(a, b) and AIF(a, b), for example, can never be true since a method can always be invoked between any pair of events a and b. 
To address these shortcoming and to retain the benefits of these properties in describing temporal constraints, 
we observe that side-effect-free method invocations can never affect the state of a software system, and as such, 
can be abstracted away in the description of the "immediately" variants of the LTL properties.
We reformulate these variants to incorporate knowledge of side-effect free methods:
\begin{enumerate}
  \item AIF(a, b): an occurrence of event a must be immediately followed by an occurrence of event b 
, ignoring all occurrence of side-effect-free events. 
  In LTL, this rule is $G(a \rightarrow X~((p_1 \lor p_2 \lor p_3 \lor \dots \lor p_n)~U~b)$, 
  where $p_1$, $p_2$,\dots,$p_n$ are side-effect free events, known ahead of time.
  \item NIF(a, b): an occurrence of event a is never immediately followed by event b, ignoring the occurrence of side-effect-free events. 
  In LTL, this rule is $G(a \rightarrow X(p_1 \lor p_2 \lor p_3 \lor \dots \lor p_n)~U~\neg b)$,
  where $p_1$, $p_2$,\dots,$p_n$ are side-effect free events, known ahead of time.
  \item AIP(a, b): an occurrence of event a must be immediately preceded by event b, ignoring the occurrence of side-effect-free events. 
  In LTL, this rule is $F(a) \rightarrow (\neg a~U (b \land  X ((p_1 \lor p_2 \lor p_3 \lor \dots \lor p_n)~U a)))$
  where $p_1$, $p_2$,\dots,$p_n$ are side-effect free events, known ahead of time.
\end{enumerate}

To identify side-effect free methods of a class, 
we use a lightweight static analysis~\cite{huang2012reiminfer} and a heuristic based on the method name (we consider names starting with "is-" or "has-" to be getters, which are typically pure).
While static analysis is used to partially accomplish this, it is not necessary for our approach. 
When neither source code nor bytecode is available, a developer can annotate the purity of relevant events by hand.

To mine LTL specifications, we use a solution similar to the linear miner algorithm described by Lemieux et al.~\cite{lemieux2015general}. 
However, as we are only interested in 2-event rules and we restrict ourselves to a few properties, 
we do not face the challenges that they solve. We use a simple way of iterating over the traces and try to use each trace to falsify the LTL specifications. 

In the work by Lemieux et al.~\cite{lemieux2015general}, support and confidence thresholds are needed. 
The support counts the number of times while iterating through the traces where a property can be falsified, but is not. 
The confidence of a property is the ratio of support of a property to the number of times the property can be falsified. 
When the number of times the property can be falsified is 0, the confidence is defined to be 1. 

For our work, as we are interested only in properties that are never falsified, thus in our implementation, 
we require a confidence of 1.0 for the rules that we mine. Moreover, we only require a support of 1 to admit the temporal property. 
In other words, as long as a property holds on a trace, and we do not encounter any trace that contradicts the property, we admit the property.

\subsection{Adversarial Test Generation}

The specification mining process adversarially generates test cases against an input specification, 
aiming to invalidate the specification.
To this end, DICE converts the temporal specifications mined from the previous phase into search goals for search-based testing. 
The objective of this phase is to allow for the discovery of test cases that produce traces that are uncommon and not represented in the initial test suite.
To do so, the temporal specifications are converted into fitness goals for test generation. 
DICE-Tester generates test cases using Evosuite\footnote{Evosuite version 1.0.6 was used in this study} with the addition of the new fitness goals and coverage criteria.  

A fitness cost is computed for each fitness goal for each test case that is generated. 
A single test case may contain multiple object instances of the class of interest and 
we consider a single trace to be the methods invoked on a single object instance. 
Therefore, each test may produce multiple traces, one for each object instance in the test case.
The fitness of a test, $T$, with respect to a fitness goal, $G$, is determined by the trace with the best fitness.
\begin{center}
  $$Fitness(T, G) = min(Fitness(tr, G)) \;\;,\;\; tr \in traces(T) $$
\end{center}

We define a new coverage criterion, \texttt{TemporalPropertyCounterExample(PropertyType, EventA, EventB)}, based on the LTL properties we have mined. 
An temporal property is covered if a trace contains a sequence of method invocations that is a counterexample of it. 
In this formulation, DICE-Tester creates a fitness goal for each temporal property we have mined, guiding Evosuite to produce counterexamples for them. 
With respect to a single fitness goal, a test is fitter than another if the fitness cost of the test is lower than the other. 
A goal is covered when the fitness cost is 0, i.e., when a counterexample trace is produced when exercising the test.

When the property is not covered, we aim to guide the test generation process towards a test covering the property.
Generally, a trace is scored relative to the number of modifications required to transform the trace to, 
first, a trace supporting the temporal property, 
then, to falsifying the temporal property.
This will push test generation towards test cases that first support the property,
from which they may be mutated towards counterexamples afterwards.

To this end, we grant a better fitness cost when the trace contains one method of a "never" property (NF, NIF) 
and when the trace contains both methods of an "always" property (AP, AIP, AF, AIF).
To summarize, for a temporal property AP(A,B) or AIP(A,B), we assign fitness costs to traces such that the following ordering holds:
\begin{enumerate} 
  \item Traces falsifying the temporal property (best)
\item Traces supporting the temporal property
\item Traces containing at least one of the events, A or B, but neither supporting or falsifying the temporal property.
\item Traces where none of A and B are present (worst)
\end{enumerate}

The above ordering guides our design of the fitness functions for all property types.
Using the above ordering, the traces are partitioned into 4 subsets for the "always" properties, AP, AIP, AF, and AIF.
In our implementation, as Evosuite expects a numerical cost, we evenly partition the range [0, 1] into the 4 subsets.
Formally, each trace, $tr$, is assigned a cost for the search goal targeting a property, $p$, relating two events, A and B, based on the following conditions:

\[
Fitness(tr, p) = \begin{cases} 
0.0 &\mbox{if falsifies(tr, p)}  \\
0.33 &\mbox{if  !falsifies(tr, p) and supports(tr, p)} \\
0.66 &\mbox{if !falsifies(tr, p) and  !supports(tr, p)} \\ 
            &\mbox{\hspace{0.25cm}and (contains(tr, A) or contains(tr, B)) }\\
1.0 &\mbox{otherwise} 
\end{cases} .
\]

To provide some intuition, we use an example using a property AP(A, B), event B must precede event A.
A trace in which event B does not precede event A, e.g. [X, Y, A], will falsify the property.
Given a trace [B, X, Y, A], which supports the property,
the trace is one transformation away, in which the method call B is removed, away from falsifying the property. 
For another example, using a trace [B, X, Y].
This trace is one edit from supporting the property. 
The event A is added in any of the three positions (between B and X, between X and Y, after Y).
After this, the modified trace is one edit from falsifying the property AP(A, B).
As it took two edits, the trace [B, X, Y] is less fit than the trace [B, X, Y, A] which only requires one edit.
Observe that a trace may containing sequence of events that both supports and falsifies a property.
A trace [A, B, A] has both the sequence [A], which falsifies AP(A,B), and [B, A], which supports AP(A,B). 
According to our fitness formulation, this trace will obtain a fitness cost of 0, since it falsifies the property.

A trace without both events A and B, e.g. [X, Y], has the worst fitness cost for the AP(A, B) search goal.
This may be contrary to intuition, since a trace without any of the property's events may resemble a trace that has a few edits away from falsifying the property.
For example, the trace [X, Y] is only 1 edit from falsifying the property directly, e.g. [X, Y] to [X, Y, A].
There are numerous such traces among the entire test population.
An objective of our fitness functions is to focus the search on properties that are hard to falsify. 
If the property can be easily covered by having such a trace transformed to a counterexample trace, 
there would be numerous such cases. 
It is very likely that 
the search goal can be covered as part of collateral coverage when the entire test population evolves to satisfy other coverage goals.
On the other hand, temporal properties that are more difficult to cover will benefit from the focused search.

For example, consider the property AP(\texttt{isEmpty:false}, \texttt{push:true}) for a hypothetical data structure, 
which holds if the data structure cannot report that it is not empty unless the push method has been successfully invoked. 
A counterexample trace can be constructed, e.g. [\texttt{<init>}, \texttt{pushAll:true}, \texttt{isEmpty:false}], by using 
an alternative method (\texttt{pushAll:true}) to insert an item into the data structure.
From a trace supporting the event, e.g. [\texttt{<init>}, \texttt{push:true}, \texttt{isEmpty:false}], a single transformation from \texttt{push:true} to \texttt{pushAll:true} 
will lead to a trace falsifying the property.
On the other hand, given an initial trace [\texttt{<init>}, \texttt{get:false}, \texttt{clearAllElements}] that does not contain either event in the property, 
if we try to falsify the property by adding a \texttt{isEmpty} method invocation, instead of adding the \texttt{isEmpty:false} event, the event \texttt{isEmpty:true} added to the trace.
It is not possible to arbitrarily add an \texttt{isEmpty:false} event into any position in a trace.
The \texttt{pushAll:true} event has to be successfully added first. 
As a result, a randomly generated trace without the \texttt{isEmpty:true} event, e.g. [\texttt{<init>}, \texttt{get:false}, \texttt{clearAllElements}], 
is much further than a trace supporting the property, e.g. [\texttt{<init>}, \texttt{push:true}, \texttt{isEmpty:false}]. 

For a temporal property AF(A,B), a trace is scored similarly. 
Observe that the fitness cost of AF(A,B) can be computed by using the same scoring rules for AP(B, A) and reversing the trace. 
Likewise, AIF(A,B) can be computed using AIP(B, A) by reversing the trace.

For a NF property, NF(A,B), the same ordering of traces described above applies.
A trace supporting the property is a trace where A is present, but B does not appear after A.
In the "immediately" variation of NF, NIF, 
the fitness functions can be computed to differentiate individual traces from one another with higher granularity, 
while still respecting the ordering of traces.
In NIF, the fitness cost reflects how closely positioned the two events in the property are within the trace.
A better fitness cost
is returned if both events are included in the trace, and we score the fitness value based on how far apart the method invocations are. 
This will push test generation towards tests where the events are located nearer to each other.
The fitness cost of a trace, $tr$, with respect to the fitness goal of the property NIF(A,B) is computed as shown in Algorithm \ref{algo:fitness}.

\begin{algorithm}
  \SetAlgoLined
  \KwIn{A trace, $tr$. A fitness goal for a NIF property, $NIF(A,B)$.}
  \KwOut{Fitness cost of the trace. 0 means that the property is covered}
  eventAPosition = -1\;
  i = 0\;
  distance = 999\;
  counter = 0\;
   \For{event $\gets$ tr} {
     \eIf{event = A}{
      eventAPosition = i\;
      counter = 0\;
     }{
      \If{$eventAPosition \neq -1$}{
          \If{!isPure(event)} {
              counter += 1\;
          }
          \If{event = B}{
              distance = Min(counter, distance)\;
          }
      }
     }
   }
   \eIf{distance = 0}{
      \Return{0} 
   }{
      \eIf{eventAPosition = -1} {
        \tcc{Event A does not appear, trace is irrelevant to this goal}

        \Return{1} 
      }{
        \Return{distance / (length(tr))}
      }
   }
   \caption{Fitness computation of a trace, tr, with respect to \texttt{TemporalPropertyCounterExample(NIF, A, B)}}
   \label{algo:fitness}
\end{algorithm}

Algorithm \ref{algo:fitness} takes a single trace, consisting of a sequence of events, 
as input and returns the fitness score of this trace. 
The fitness score ranges from 0 to 1, 
and a score of 0 indicates the maximum fitness value while a score of 1 indicates that the trace has no relevance for this goal.
A single pass is made through the sequence of events to look for instances of event A (lines 5-19).
A counter (line 4) to measure the distance between a pair of events A and B is updated in this pass. 
Each time we reach an instance of event A, we record its position and reset the counter (lines 7,8).
Once an instance of event A has been found (line 10), 
we increase the counter for each impure (i.e. not side-effect free) event we pass (line 11-12).
Whenever we reach an instance of event B, 
we update the minimum distance between the last seen instance of A and B if the counter is 
less than the previous minimum distance between A and B (line 14-16).
Finally, the fitness score is computed from the minimum distance. 
If the distance is 0, then the goal is covered and 0 is output as the fitness score (line 20-22).
Otherwise, the ratio of the distance to the length of trace is used as the fitness score (line 23-27). 

To conclude, 
the satisfaction criterion of each \textit{TemporalPropertyCounterExample} depends on the property type.
Effectively, if a test contains a trace with a fitness score of 0, then the test is a counterexample.
It may be interpreted as follows:
\begin{itemize}
  \item AF(a,b): a test is a counterexample of AF(a, b) if it contains a trace with an invocation of a that is not (immediately) followed by b. 
  \item NF(a,b): a test is a counterexample of NF(a, b) if it contains a trace with an invocation of a that is (immediately) followed by b. 
  \item AP(a,b): a test is a counterexample of AP(a, b) if it contains a trace with an invocation of b that is not (immediately) preceded by a. 
\end{itemize}

As a consequence of including our new fitness goals into Evosuite, 
there are a large number of search objectives (one for each temporal property we have mined). 
As described in Section \ref{sec:background}, 
this hampers the ability of Evosuite to search effectively for good test cases that will cover the uncovered goals. 
The large number of search goals causes the search process to be similar to a random search process, 
reducing its performance. 
To manage the large set of goals, we use the DynaMOSA search algorithm as discussed earlier in Section 2. 
For each search goal, the DynaMOSA algorithm uses a test ranking strategy which ensures that the test case closest to covering it always survives to the next generation 
when tests are ranked and weaker tests are discarded.
Furthermore, it models the dependencies between structural goals.
At any time, only the goals with fully satisfied dependencies are in the current set of goals.
This strategy prevents goals that cannot contribute meaningfully to the test ranking from becoming targets.  
However, we found that the existing implementation of the DynaMOSA search algorithm was not sufficient to enable Evosuite to find counterexamples effectively.

  Thus, we use the DynaMOSA search algorithm in DICE-Tester with three modifications.
  To further boost the efficiency of selecting tests, our first modification is to model the dependencies between the coverage goals beyond structural goals to allow Evosuite to dynamically select goals more effectively. 
  While the DynaMOSA algorithm already models the dependency between structural coverage goals based on control dependencies, 
  it does not model other forms of dependencies, 
  including usage dependencies between methods. 
  It is unable to perform any reasoning about method coverage goals, 
  which DICE-Tester is able to constrain using the LTL properties we have mined previously.
  
  The dependency tree is constructed prior to the start of the search process.
  At this stage, we assume that every mined LTL properties is true.
  This dependency tree relates method coverage goals with the \texttt{TemporalPropertyCounterExample} coverage goals that we introduced in this work. 
  A method coverage goal, \texttt{Method(A)}, is covered when a test case invokes the method A at least once. 
  We add dependencies based on the following rules:
  \begin{itemize}
      \item Given AP(A, B), DICE-Tester adds a dependency where B depends on A. i.e. \texttt{Method(A)} should be covered before \texttt{Method(B)}
      \item Given NF(A, B) or NIF(A,B), DICE-Tester adds a dependency where \texttt{Method(A)} should be covered before \texttt{TemporalPropertyCounterExample(NF, A, B)} or \\
      \texttt{TemporalPropertyCounterExample(NIF, A, B)}.
      
      \item Given AF(A, B) or AIF(A,B), DICE-Tester adds a dependency where \texttt{Method(A)}should be covered before \texttt{TemporalPropertyCounterExample(AF, A, B)} or \\
      \texttt{TemporalPropertyCounterExample(AIF, A, B)}.
      
      \item Given AP(A, B) or AIP(A,B), DICE-Tester adds a dependency where \texttt{Method(B)} should be covered before \texttt{TemporalPropertyCounterExample(AP, A, B)} or \\
      \texttt{TemporalPropertyCounterExample(AIP, A, B)}.
      
  \end{itemize}
  
  For example, given \texttt{NF(StringTokenizer, nextToken)}, for its corresponding goal \\
  \texttt{TemporalPropertyCounterExample(NF, StringTokenizer, nextToken)} to be satisfied, 
  the method coverage goal, \texttt{Method(StringTokenizer)}, must be satisfied first. 
  If this method coverage goal is not satisfied, then it is impossible for the test to produce a counterexample and this goal is always completely uncovered for any test case. 
  Thus, it will not contribute meaningfully to the ranking of tests maintained by DynaMOSA~\cite{fraser2011evosuite}. 
  Constraints about the ordering of methods provided by the set of AP temporal properties also allow us to add dependencies between a pair of method coverage goals. 
  For another example, given \texttt{AP(initVerify,update)}, we add a dependency between \texttt{Method(update)} and \texttt{Method(initVerify)}, 
  requiring an invocation of \textit{initVerify} before Evosuite adds \textit{update} to the set of currently targeted goals.
  While other coverage goals may return a fitness value between 0 and 1, 
  the fitness of a test with respect to a method coverage goal is binary, either the goal is covered or it is not. 
  Therefore, if the prerequisites are not met, a method coverage goal is always uncovered for every test case.

  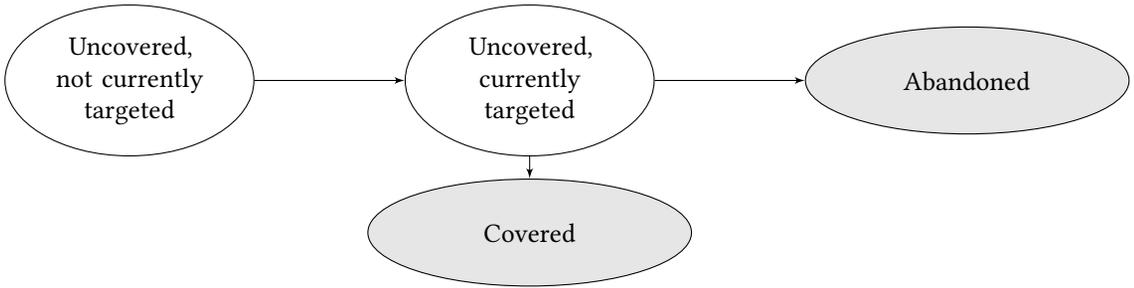
\begin{figure*}
    \begin{center}
      
    \begin{tikzpicture}[node distance = 2.0cm , yscale=1.3,
      ,line/.style={draw, -latex'},
      test/.style=%
        {%
          ellipse, draw, text width=6em, text centered, rounded
          corners, minimum height=4em
        },
        trace/.style=%
        {%
        rectangle, draw, text width=8em, text centered, rounded
          corners, minimum height=4em, minimum width=4em
        },
        fsa/.style=%
        {%
        ellipse, draw, fill=gray!20,  text width=8em, text centered, rounded
          corners, minimum height=4em, minimum width=4em
        },
        outer/.style={draw=gray,dashed,fill=green!1,thick,inner sep=5pt },
        miner/.style=%
        {%
        rectangle, draw,   text width=8em, text centered, minimum height=4em, minimum width=4em
        }]
        \node [test] (uncovered_not_current) {Uncovered, not currently targeted};
        \node [test,right= of uncovered_not_current] (uncovered_current) {Uncovered, currently targeted};
          \node [fsa,below of=uncovered_current] (covered) {Covered};
  
        \node [fsa, right=of uncovered_current] (abandoned) {Abandoned};

        \path [line] (uncovered_not_current) -- (uncovered_current) ;
        \path [line] (uncovered_current) -- (covered) ;
        \path [line] (uncovered_current) -- (abandoned) ;

    \end{tikzpicture}
      
    \end{center}
    \caption{Lifecycle of a search goal. At the end of the search process, a goal is either covered or abandoned.}
    \label{fig:goal_lifecycle}
  \end{figure*}
  
  Our second modification is made for the lifecycle of a search goal.
  In the DynaMOSA algorithm, goals are in one of 3 states. 
  A goal is either a) covered, b) uncovered but not in the current set of goals, or is c) uncovered but in the current set of goals. 
  In DICE-Tester, we added a new state in this lifecycle. 
  Thus, a goal can be in one of 4 states: covered, uncovered but in the current set of goals, uncovered but not in the current set of goals, or \textit{abandoned} (see Figure \ref{fig:goal_lifecycle}).
  We refer to a goal as \textit{abandoned} if it is not covered but we have removed it from the current set of goals.
  
  The addition of this state is motivated by the fact that at least some of the mined properties are true. 
  As these properties are true, it is impossible to generate a test case that is a counterexample to it. 
  Consequently, goals representing counterexamples to these properties can never be covered, and this has detrimental effects on the search process. 
  Such goals will continue to contribute to the preference ranking used in Evosuite. 
  This has several implications. 
  Test cases that are closest to and "almost covering" the goal will always remain in our population, although they will not contribute to producing interesting test cases. 
  The search may be weighted towards tests that have a higher chance of covering these goals, even though these goals cannot be covered. 
  This may potentially prevent other test cases (that may lead to a counterexample of another property) from getting added to the population. 
  Furthermore, these goals have no meaningful contribution to the ranking process when weighing the pareto-optimality of the other test cases. 
  This slows down the search process, wasting time to compute the fitness of tests with respect to these goals.
  
  Hence, we add the abandoned state in the lifecycle of a goal. 
  To enable goals to transit to this state, DICE-Tester tracks the age of each goal. 
  Covered goals and goals that are not in the set of currently targeted goals do not have an age. 
  When an uncovered goal moves into the set of targeted goals, its age is initialized to 0. 
  We increment the age of a goal for each generation where DICE-Tester does not find a test case that covers it. 
  Once the age of a goal has exceeded a threshold, we abandon the goal by removing it from the set of targets. 
  In our experiments, we set this threshold to 100 generations. 
  Once abandoned, goals can never be restored back to the set of targets. 
  
  Our third modification is to reset the population of the tests once it has gotten stuck.
  While the first two modifications allow Evosuite to find more counterexamples, 
  we observed that the search can still get stuck in a local optima. 
  Finally, we bypass this problem by resetting the population of tests, and in effect restarting the search, 
  once 100 generations has passed without DICE-Tester finding a test that is a counterexample to \textit{any} goal. 
  We did not thoroughly empirically evaluate the threshold for abandoning a goal or resetting the population, 
  however, we noticed in our experiments that changing these parameters do not affect the results much, 
  provided that they are not too small.
  
  With these three modifications, DICE-Tester is able to guide test generation 
  towards finding counterexamples to spuriously mined LTL properties, to invalidate them. 
  Temporal properties with at least one counterexample are removed. 

  Next, we collect execution traces from the output test suite, which are constructed from the test cases that produced counterexamples to the temporal properties. 
  These are later passed as input to the FSA inference process.
  Typically, specification mining algorithms use traces of correct executions
  and our approach is not an exception.
  Therefore, we filter out traces of executions that may not represent a correct usage. 
  If the invocation of a method results in a thrown exception, 
  we ignore the invocation  and any further method invocations (as the exceptional invocation may have an effect on the state). 
  Apart from exceptional executions,
  we also try to detect resource leaks and omit any possible trace that caused them. 
  We keep track of the number of file descriptors that are opened by the process that is executing the tests~\footnote{We run Evosuite without its sandbox which prevents environmental interactions}.
  Next, we compared the number of file descriptors before and after the runs, 
  and if we find a mismatch between  them, we assume that the tests have triggered a sequence of methods that leaked a file or a socket.
  When this happens, we remove all traces of tests that led to resource leaks. 
  While this may inadvertently result in correct traces that are incorrectly discarded 
  (consider a scenario where a test instantiated multiple FileOutputStreams, 
  we discard traces from every FileOutputStream as long as one of them caused a leak, 
  even if the rest of the instances represent correct usage of FileOutputStream), 
  this helps us ensure that the traces we have collected do not contain executions of invalid usage of an API or class.

\subsection{Example of the search process}

\begin{figure}[h]
	\centering
	\scriptsize{
\begin{lstlisting}[language=java,numbers=none,escapechar=!,basicstyle=\ttfamily]
interface DataStructure<T> {

    public boolean add(T item); 

    public boolean addAll(Collection<T> items);

    public boolean isEmpty();

    public void clear();

    public T get();

    public Collection<T> getAll();
}
\end{lstlisting}
    \caption{The API of a hypothetical data-structure.}
    \label{fig:data_structure}
	}
\end{figure}

In this section, we present a synthetic example of the search process. 
Using a hypothetical data-structure shown in Figure \ref{fig:data_structure},
we show how a small set of test cases may evolve over a few generations to become counterexamples for 
a three properties that are not true properties:
\begin{itemize}
  \item Goal 1: AP(isEmpty:FALSE, add:true),
  \item Goal 2: AIF(clear, isEmpty:TRUE), and
  \item Goal 3: AF(clear, isEmpty:TRUE), 
\end{itemize}

We run the DICE process to search for test cases that will falsify the properties.
In this simple example, all three properties are falsifiable.
As described earlier, we use the dependencies between search goals 
to consider fewer search goals at a time, omitting goals that cannot be covered yet.
At the start of the DICE-Tester process, 
the following dependencies between search goals are added: \\

$Method(isEmpty:FALSE) \rightarrow {AP(isEmpty:FALSE, add:true)} $

$Method(isEmpty:TRUE) \rightarrow {AIF(clear, isEmpty:TRUE),AF(clear, isEmpty:TRUE)}$

$Method(clear) \rightarrow {AIF(clear, isEmpty:TRUE),AF(clear, isEmpty:TRUE)}$

$Method(add:TRUE) \rightarrow {AP(isEmpty:FALSE, add:true)}$ \\

The dependencies should be interpreted such that the search goals on the right (the consequent) are added to the currently targeted set of goals once the search goal on the left (the antecedent) has been covered. 
The test generation process starts by creating a population of randomly generated tests.
In this example, we assume that the population size is two, and that each test produces only one trace.
First, DICE-Tester generates 2 tests, giving us the following traces as shown below in Figure \ref{fig:gen1}:

\begin{figure}[h]
	\centering
	\scriptsize{
\begin{lstlisting}[numbers=none,escapechar=!,basicstyle=\ttfamily]
  [clear, isEmpty:TRUE, getAll]                              (-, 0.33, 0.33)
  [getAll, add:FALSE]                                        (-, 1.00, 1.00)
\end{lstlisting}
    \caption{The initial population. The numbers on the right show their fitness costs for Goals 2 and 3. Goal 1 is not considered yet as neither of the method coverage goals it depends on has been covered.}
    \label{fig:gen1}
	}
\end{figure}

In Figure \ref{fig:gen1}, Goal 1 is not considered among the search goals (listed as '-') as neither of the method coverage goals it depends on has been covered.
When the number of goals is large, this helps to simplify the comparison between test cases.
When this initial population of test cases is evolved, it produces a set of new test cases as shown in Figure \ref{fig:gen1_offspring}. 
Since the method coverage goal of Goal 1 has been covered, all the search goals for the 3 properties are now considered.

\begin{figure}[h]
	\centering
	\scriptsize{
\begin{lstlisting}[numbers=none,escapechar=!,basicstyle=\ttfamily]
  [clear, getAll, isEmpty:TRUE, getAll]                      (1.00, 0.00, 0.33)   *
  [clear, isEmpty:TRUE, add:TRUE, isEmpty:FALSE]             (0.33, 0.33, 0.33)   *
  [add:TRUE, clear, isEmpty:TRUE, getAll]                    (0.66, 0.33, 0.33)
  [getAll, isEmpty:TRUE, add:TRUE, getAll]                   (0.66, 0.66, 0.66)
  [getAll, add:TRUE, get]                                    (0.66, 1.00, 1.00)
\end{lstlisting}
    \caption{The offspring of the first generation. We order them by their pareto-optimality.}
    \label{fig:gen1_offspring}
	}
\end{figure}

In Figure \ref{fig:gen1_offspring}, the test cases that are selected to form the next population of tests are indicated with a *.
As the first test case has fully covered Goal 2 (falsifying the property that \texttt{clear} is always followed by \texttt{isEmpty:TRUE}),
it is selected. 
As for the second test case, it ties with the third test case for the best fitness score on Goal 3, 
but it is the best test case for Goal 1.
Therefore, both the first and second test cases are selected as they are the best test cases with respect to the search goals.
The remaining test cases are ordered by pareto-optimality. 
The third test case dominates the fourth test case, and the fourth test case dominates the fifth test case.
The first test case is inserted into the archive, as it has fully covered a search goal.
If the population size was greater than 2, then the remaining test cases would be added in this order.

Finally, in the next round, we get the following offspring shown in Figure \ref{fig:gen2_offspring}.
As search goal 2 was already fully covered earlier, we no longer need to consider
the test cases' fitness with respect to it.

\begin{figure}[h]
	\centering
	\scriptsize{
\begin{lstlisting}[numbers=none,escapechar=!,basicstyle=\ttfamily] 
  [clear, isEmpty:TRUE, addAll:TRUE, isEmpty:FALSE]         (0.00, -, 0.33)   *
  [clear, clear, getAll]                                    (0.66, -, 0.00)   * 
  [clear, getAll, isEmpty:TRUE, add:TRUE, getAll ]          (0.66, -, 0.66)
  [clear, isEmpty:TRUE, sEmpty:TRUE, add:TRUE, getAll]      (0.66, -, 1.00)
\end{lstlisting}
    \caption{The offspring of the second generation of test cases. All the test cases are covered after this. Note that Goal 2 was not considered when computing the objective function vectors as it has a corresponding test case in the archive.}
    \label{fig:gen2_offspring}
	}
\end{figure}

Both Goals 1 and 3 have counterexample traces from this set of offspring, 
and these two test cases are added to the archive.
As all the search goals related to the temporal properties are covered, we end the test generation process.
In practice, it is typically the case that not all coverage goals can be covered and the test generation process 
only ends when the search budget is fully consumed.
The test cases from the archive are retrieved to form the output test suite, 
which is run and its execution traces are collected to be used as input to the next step of DICE.

\subsection{FSA Inference}
\label{sec:approach_first_step}

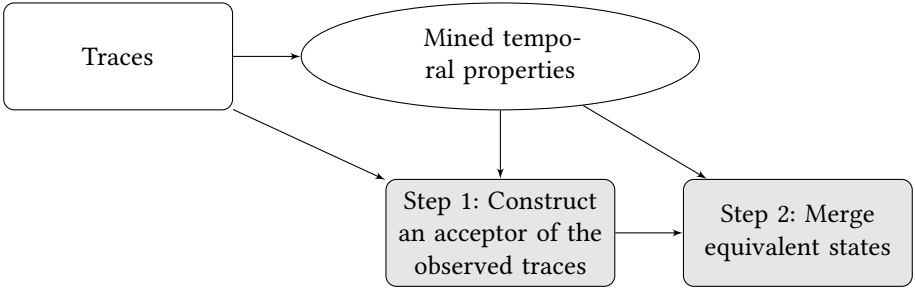
\begin{figure*}
  \begin{center}
    
  \begin{tikzpicture}[node distance = 0.9cm , yscale=0.5,
    ,line/.style={draw, -latex'},
    test/.style=%
      {%
        ellipse, draw, text width=10em, text centered, rounded
        corners, minimum height=4em, minimum width=4em
      },
      trace/.style=%
      {%
      rectangle, draw, text width=8em, text centered, rounded
        corners, minimum height=4em, minimum width=4em
      },
      fsa/.style=%
      {%
      rectangle, draw, fill=gray!20,  text width=8em, text centered, rounded
        corners, minimum height=4em, minimum width=4em
      },
      outer/.style={draw=gray,dashed,fill=green!1,thick,inner sep=5pt },
      miner/.style=%
      {%
      rectangle, draw,   text width=8em, text centered, minimum height=4em, minimum width=4em
      }]
      \node [trace] (trace) {Traces};
      \node [test,right= of trace] (temporal_properties) {Mined temporal properties};
      \node [fsa,below = of temporal_properties] (acceptor) {Step 1: Construct an acceptor of the observed traces};
      \node [fsa, right=of acceptor] (merging) {Step 2: Merge equivalent states};

      \path [line] (trace) -- (temporal_properties) ;
      \path [line] (trace) -- (acceptor) ;
      \path [line] (temporal_properties) -- (acceptor) ;
      \path [line] (acceptor)  -- (merging);
      \path [line] (temporal_properties) -- (merging);

  \end{tikzpicture}
    
  \end{center}
  \caption{High-level overview of DICE-Miner}
  \label{fig:spec_mining_approach}
\end{figure*}

The final phase of the adversarial specification mining approach is to 
take the temporal properties and traces produced from the previous phases
and infer an FSA model.
In DICE, the DICE-Miner algorithm is responsible for this. 
We add the traces produced by the DICE-Tester to the original set of traces for input to DICE-Miner.

An overview of DICE-Miner is given in Figure \ref{fig:spec_mining_approach}.
The DICE-Miner algorithm to infer an FSA-based specification is similar to the k-tails algorithm, comprising of two steps.
In the first step, we construct an initial automaton based on the input traces and the mined temporal properties. 
Similar to the PTA (described earlier in Section \ref{sec:background}), 
our initial automaton accepts all of the observed input traces. 
The mined temporal properties are used to prevent erroneous merging of states. 
In the second step, we merge equivalent states in this automaton, leveraging the mined temporal properties. 
The mined temporal properties are used to derive the set of enabled methods of each state, 
which is the equivalence criteria used in DICE-Miner to merge states. 
We elaborate on this equivalence criteria below in Section \ref{sec:spec_second_step}.

Regarding the first step of the algorithm, we make the following observations:
\begin{itemize}
  \item Observation A: Side-effect-free events can be interleaved in any order, and do not change the present state of the software system. 
  \item Observation B: The first step of constructing a PTA makes the assumption that states with the same history of events are equivalent. This assumption may not be true.
\end{itemize}
We propose a specification mining algorithm with these observations in mind. 
In light of the first observation, 
we include information of the purity of each method to allow the model to have greater generalizability.
The second observation guided us to prevent the incorrect conflation of states while constructing an initial model that accepts all of the input traces.

With observation A, we model the freedom of side-effects as self-loops in the automaton. 
We add Constraint A where we ensure that transitions labelled with side-effect free method calls are self-loop.
This has the advantage of increasing the model's generalizability, 
allowing it to accept an equivalent trace with a different permutation of the pure methods. 
For example, for \texttt{StringTokenizer}, 
observing an input trace [\texttt{StringTokenizer}, \textbf{\texttt{hasMoreElements:true}}, \texttt{hasMoreTokens:true}]  
allows the construction of an automaton that will 
accept a different trace (note a different ordering of the method invocations), [\texttt{StringTokenizer}, \texttt{hasMoreTokens:true}, \textbf{\texttt{hasMoreElements:true}}].
This is the case even without the observation of a trace with this sequence of method calls, since the only difference is that the 
pure methods \textit{hasMoreTokens} and \textit{hasMoreElements} were invoked in a different order from the same state.

To address observation B, we add Constraint B where we prevent the erroneous conflation of states occurring in the first step of k-tails and its variants.  
If these states are incorrectly merged in the first step, 
regardless of the equivalence criteria selected for merging states, 
this inaccuracy in the initial model will negatively impact the quality of the final FSA produced.
This is because the second step does not split up incorrectly merged states from the first step.
For an example of Observation B, with the \texttt{Iterator} class, given the trace 
  [\textbf{\texttt{Iterator}, \texttt{hasNext:true}, \texttt{next}}, \texttt{hasNext:false}]  and 
  a different trace sharing a prefix, ~[\textbf{\texttt{Iterator}, \texttt{hasNext:true}, \texttt{next}}, \texttt{hasNext:true}],
the states reached in the two traces after the invocation of \textbf{next} is different, yet constructing a PTA will conflate these states as they share the same preceding events [\texttt{Iterator}, \texttt{hasNext:true}, \texttt{next}].
Conflating these states produces an automaton where both \texttt{hasNext:false} and \texttt{hasNext:true} are incorrectly enabled from the same state.

To address this, we propose to detect incorrectly merged states 
using the mined temporal properties while constructing an initial automaton accepting all of the observed traces. 
Whenever adding a new transition to an automaton results in a automaton that may produce traces violating the mined properties, 
we modify the automaton such that these violations will not occur.
Next, we discuss the details of the DICE-Miner algorithm, 
and show we address both observations within the first step of our algorithm.

\begin{algorithm}[]
\SetAlgoLined
\KwIn{Input traces, $traces$}
\KwOut{A DFA that accepts all traces in $traces$, without creating any states that may violate any constraint}
 automaton = emptyAutomaton()\;
 \For{trace $\gets$ $traces$}{
  prefix = []\;
  state = automaton.initialState\;
  \For{event $\gets$ trace}{
    \eIf{state.canAccept(event)}{
        nextState = state.accept(event)\;
        state = nextState\;
    }{
        \eIf{!state.HasIncompatibleTransition(event)}{
            newState = state.addTransitionToNewState(event)\;
            state = newState\;
        }{
            stateToModify, suffix = FindAncestorWithoutIncompatibleTransition(state, event)\;
            \For{suffixEvent $\gets$ suffix}{
                newState = stateToModify.addTransitionToNewState(suffixEvent)\;
                stateToModify = newState\;
            }
        }
        
    }
  }
  \Return{automaton}
 }
 \caption{Pseudocode for CreateCompatibleAcceptor. This is a simplified version of the algorithm. In reality, the automaton is non-deterministic and given an event, there may be multiple transitions labeled with the same event.}
 \label{algorithm:create_compat}
\end{algorithm}

\subsubsection{First Step}

Next, we describe the first step of the DICE-Miner algorithm. 
In the first step, we pass the example traces into the function \textit{CreateCompatibleAcceptor}, which constructs a Non-deterministic Finite Automaton (NFA).
This automaton is build to accept all of the example traces, much like a PTA. 
However, unlike the construction of a PTA, \textit{CreateCompatibleAcceptor} avoids the creation of states that may accept sequences of events that violates a constraint
(we refer to such states as \textit{incompatible} with the constraint). 
We consider a state to violate a constraint if it causes the automaton to accept a trace violating the constraint.
Observe that although the states were constructed based on individually observed traces, and that every trace does not violate 
the temporal properties mined earlier,
it is possible to construct a PTA with states that may accept traces violating the temporal properties.
For example, 
given two traces, [\texttt{Stack}, \texttt{addAll}, \texttt{remove}, \texttt{isEmpty:TRUE}] and [\texttt{Stack}, \texttt{addAll}, \texttt{remove}, \texttt{get}] and 
a temporal property, \textit{NIF(isEmpty=True, get))}, 
a state with an incoming edge, \texttt{isEmpty:TRUE} (a self-loop on the state, as a result of Constraint A and the fact that \texttt{isEmpty} is pure), 
and an outgoing edge, \texttt{get}, is \textit{incompatible} with the temporal property as it can accept a trace with the sequence of events [\texttt{isEmpty:TRUE}, \texttt{get}].
The algorithm to split up a state with incompatible edges is given in Algorithm \ref{algorithm:create_compat}.

\textit{CreateCompatibleAcceptor} first initializes an empty automaton before iterating over each trace (lines 1-2). 
For each event in a trace, we first try to accept the event without modifying the NFA (lines 5-8). 
On reaching an event, $e$, that cannot be accepted, we modify the NFA to add new states and transitions such that it will accept the events. 
Before adding a new state, we first ensure that the new transition will not cause the current state to be incompatible with any constraint (line 10).
If adding the transition results in an incompatible state (lines 13 - 19), we traverse backwards, looking for an ancestor where we can add transitions corresponding to the events up to $e$ without introducing an incompatible state (line 13). 
From this ancestor node, we add new transitions and states to represent the events up to $e$ (lines 15-18).

\begin{algorithm}[]
\SetAlgoLined
\KwIn{The current state, $state$, and the event that introduced an incompatible state, $event$}
\KwOut{An ancestor state and a sequence of events to add transitions for}
 parent = state\;
 suffix = []\;
 label = event\;
 \While{parent.HasIncompatibleTransition(label)} {
    grandParent, label = parent.getIncomingTransition()\;
    suffix = label :: suffix\;
    parent = grandParent\;
 }
 \Return{parent, suffix}
 
 \caption{Pseudocode of FindAncestorWithoutIncompatibleTransition. Traverses the ancestry of a state to locate a state to branch from.}
 \label{algorithm:find_ancestor}
\end{algorithm}

To find an ancestor from which it is possible to add a new chain of events such that an event can be added without incompatibility, 
we use the algorithm \textit{FindAncestorWithoutIncompatibleTransition}. 
The details of \textit{FindAncestorWithoutIncompatibleTransition} are given in Algorithm \ref{algorithm:find_ancestor}. 
We initialize the algorithm (lines 1-3) before 
we iteratively traverse the ancestors of the parent state. 
At this state of the DICE-Miner algorithm, all states have at most 
one incoming transition originating from another state, which is obtained using \texttt{getIncomingTransition}.
As we traverse backwards, we collect the labels on the transitions (line 6).
These labels correspond to the events that we will have to add transitions for. 
The traversal ends once we find an ancestor that we add the sequence of events including $e$ without creating an incompatible state.

\begin{algorithm}[t]
\SetAlgoLined
\KwIn{A state, $state$, and an event to add, $event$}
\KwOut{true if adding the event does result in the state becoming incompatible}
 \For{transition $ \gets $ state.incomingTransitions} {
   \If{NIF(transition,event)}{
    \Return{true}
   }
 }
 \eIf{NF(state.prefix,event)}{
    \Return{true}
  }{
    \Return{false}
 }
 
 \caption{Pseudocode of HasIncompatibleTransition. Checks if a transition labeled with the event can be added to the input state. }
 \label{algorithm:incompatible_transition}
\end{algorithm}

Algorithm \ref{algorithm:incompatible_transition} shows the check for an incompatible state. 
As we only add new outgoing transitions, it is sufficient to check pairs of the existing incoming transitions with the new event to detect NIF violations (lines 1-5).
To check NF violations, we check the prefix of the state against the event (line 6). 
The prefix of each state is constructed by traversing all transitions on the trail of ancestor states. 
This includes all self-loops on each ancestor (i.e. the self-loops are traversed first before moving to a child state).
It is enough to check for violations of NF and NIF, as the other properties are never violated in the first stage of DICE-Miner.

At the end of the first step, we receive an automaton that contains self-loops in some states, and these are the only cycles in it. 
The inferred model is a Non-Deterministic Automaton (NFA) as states can have multiple transitions with the same label. 
We presented simplified versions of the algorithms for ease for explanation. 
As the automata involved are NFAs, there may be more than one transition from a state given an event. 
The model is constructed based on the observed traces. 
It is sound with respect to these traces, and will accept all of these concrete traces.

\subsubsection{Second Step}
\label{sec:spec_second_step}

In the second step, we merge equivalent states in the automaton.
To merge two states, $a$ and $b$, we remove both states from the state machine, then add a new state, $c$.
All the transitions from $a$ and $b$ are added onto state $c$.
If there are multiple transitions with the same label, source, and destination, only one of them is kept and the duplicates are removed. 
As we already noted earlier, our model is non-deterministic and there may be multiple transitions from a state labelled with the same 
event.

To determine if two states are equivalent, 
we draw inspiration from the CONTRACTOR model~\cite{de2010automated} and 
define the equivalence of states based on the set of methods that are enabled. 
Concrete states that have the same enabled methods are merged. 
As described by de Caso et al.~\cite{de2010automated}, this results in models with states that are intuitively interpreted 
and are at an abstraction level that developers find convenient.  
While we reuse the concept of the enabledness of methods in order to group states, 
our method is still primarily based on the execution traces that are input to DICE.
The CONTRACTOR method requires further annotation and the computing of dependencies between the enabledness of all pairs of methods. 
In our work, we do not use these dependencies and other related concepts described by de Caso et al.~\cite{de2010automated} to 
avoid the need to annotate the pre- and post-conditions of every method.
Prior work~\cite{krka2014automatic} has also shown that the performance of the CONTRACTOR approach is highly dependent on the quality of the pre- and post-conditions, 
and that it exhausts the running memory when provided with noisy invariants. 
Instead, we use the temporal properties to aid in determining the enabledness of a method at a particular state. 
We leave the study of the applicability of the dependencies of method enabledness on DICE and their integration into DICE for future work.

However, unlike de Caso et al.~\cite{de2010automated} and Krka et al.~\cite{krka2014automatic} 
which creates models from the state invariants of an object,  
we derive the enabledness of a method from the set of NF and NIF properties. 
If there are no LTL property that \textit{disables} a method based on its prefix, the known incoming transitions, 
and the known outgoing transitions, 
then we consider that this method is enabled. 
We consider a method, say method $A$, to be \textit{disabled} on a state from following conditions. 

\[
disabled(state, A) = \begin{cases} 
true &\mbox{if } NF(X, A), X \in prefix(state) \\
true &\mbox{if } NIF(X,A), X \in incoming(state) \\
true &\mbox{if } NIF(A,Y), A \mbox{ is pure}, Y \in outgoing(state) \\
false & \mbox{otherwise}
\end{cases} .
\]

If a method on a state leads to the automaton accepting a trace containing a pair of successive events violating the property,
then the method is disabled.
For example, a state with an incoming transition \texttt{add:true} (indicating that an item was successfully added) 
will have the event \texttt{isEmpty:false} disabled (the state can't be empty after a successful addition) by the second condition. 
As another example, a state with the outgoing transition \texttt{remove:true} (indicating that an item can be successfully removed from this state) 
will have the event \texttt{isEmpty:true} disabled (if an item can be successfully removed, it means it can't be empty) by the third condition. 
Methods can be enabled even if we have not seen the method invoked from a particular state in a concrete trace. 
Before merging two states, DICE-Miner checks that it does not lead to a violation of known LTL properties. 
Similar to the work by Lo et al.~\cite{lo2009automatic}, 
a pair of states cannot be merged if it results in a state machine that violates any temporal property known to hold on the observed traces.

\section{Evaluation}
\label{sec:evaluation}

To empirically evaluate our tool, we investigate 3 research questions. 

\begin{itemize}
  \item \textbf{RQ1: How effective is DICE in inferring FSA models?} \\
  RQ1 investigates the effectiveness of the adversarial specification mining approach by comparing the FSA models inferred by DICE against models 
  inferred by state-of-the-art specification miners.

  \item \textbf{RQ2: How effective is DICE-Tester?} \\
  RQ2 investigates the effectiveness of the test generation component, DICE-Tester. 
  This is done by comparing the tool against Evosuite, with DynaMOSA as its search algorithm, 
  as a test generation baseline to answer this question.
  Instead of using the traces produced by DICE-Tester, we study if using traces produced by Evosuite is enough to  
  mine better specifications.
  Here, our objective is to evaluate the value of the temporal-property guided adversarial test generation strategy 
  we built on top of Evosuite.  

  \item \textbf{RQ3: How effective is DICE-Miner?} \\
  RQ3 investigates if the specification miner, DICE-Miner, can utilize the traces generated by DICE-Tester effectively. 
  We compare DICE-Miner against DSM as a baseline, passing the same set of traces produced by DICE-Tester as input to both tools.

\end{itemize}

\subsection{Experimental setup}

\subsubsection{Evaluation}
To investigate these questions, we empirically evaluate the tools by assessing the quality of the models inferred against ground-truth models. 
Similar to previous studies, we measure precision and recall.
This procedure for computing precision and recall has been used in prior studies~\cite{le2015beyond,krka2014automatic}. 
As input, a ground-truth model and the model inferred by DICE is provided. 
From these two models, traces are generated by randomly traversing the edges in the model. 
The precision of the inferred model is the percentage of traces produced by the inferred model that are accepted by the ground-truth model. 
The recall of the inferred model is the percentage of traces it accepts among the traces produced by the ground-truth model. 
In other words, precision is the proportion of traces  from the inferred that are correct, and recall is the proportion of correct traces that the inferred model accepts. 
Finally,
the quality of the model is measured using F-measure, computed as follows. 
\begin{center}
  $\text {F-Measure}=2 \times \frac{\text { Precision } \times \text {Recall}}{\text {Precision}+\text {Recall}}$
\end{center}

The 11 ground-truth models publicly released from Le et al.'s evaluation of DSM~\cite{le2018deep} are used~\footnote{\url{https://github.com/lebuitienduy/DSM}}. 
These evaluation library classes, used for evaluating specification miners in previous studies, 
represent 100 analysed methods in total, and represent different categories of libraries ranging from data streaming to message exchange.
To analyse the classes from JDK, we copied the source files of the corresponding classes from OpenJDK 1.8, to get around a constraint in Evosuite  
that prevented instrumentation and bytecode-rewriting of classes from some packages provided by the JDK. 
OpenJDK 1.8 was used as this was the version used in previous studies, 
and we observe that the choice of a more recent version will not impact the evaluation results 
as the ground-truth models involved only methods from earlier JDK versions. 
We also omitted traces from DICE-Tester containing events that are not present in the ground-truth models. 
This is done to allow direct comparison against the approaches used in previous studies. 
During our evaluation, we discovered a minor inaccuracy in the ground-truth model of ZipOutputStream, 
in which DICE-Tester found counterexamples to. 
Hence, we corrected the ground-truth model to account for the missing transition.
Finally, for each case,
we account for randomness by computing the average of the evaluation metrics from 20 runs of the experiment.

\subsection{RQ1}

To determine the effectiveness of our tool for answering RQ1, we compute precision, recall and F-measure of the output FSAs for 11 target library classes. 
We compare against DSM~\cite{le2018deep}, which uses deep-learning and randomized test generation~\cite{pacheco2007randoop}, as a baseline. 
For a second baseline, we also compare our tool against Tautoko~\cite{dallmeier2010generating}, 
which leverages test generation to complete an initial FSA model. 
Tautoko takes the specifications inferred by the specification miner, ADABU~\cite{dallmeier2006mining}. 
Given an initial test suite, it learns an initial model using ADABU 
and then mutates test cases and executes them again to cover missing transitions in the initial model.

\begin{table}
  \caption{Precision, Recall, and F measure of DICE and DSM. NFST refers to NumberFormatStringTokenizer.}
    \label{tab:fsm_dsm_max}
    \begin{tabular}{ |c | ccc | ccc|}
      \hline
      \multicolumn{1}{|c|}{\multirow{2}{*}{\textbf{Class}}}  &  \multicolumn{3}{c|}{\textbf{DICE}} & \multicolumn{3}{c|}{\textbf{DSM  }} \\ 
\cline{2-7} 
\multicolumn{1}{|c|}{}                      
 & \textbf{P} & \textbf{R} & \textbf{F}  & \textbf{P} & \textbf{R} & \textbf{F}        \\ \hline
      ArrayList       &  31.4  & \textbf{27.3}  &  \textbf{29.2} & \textbf{44.5} & 16.3 & 23.9 \\
      HashMap         &  100.0  &   \textbf{94.1} & \textbf{97.0} & 100.0 & 55.2 & 71.1\\
      HashSet         &  \textbf{87.4}  & \textbf{100.0}   & \textbf{93.3} & 74.0 & 62.4 &  67.7\\
      Hashtable       &  84.0  & \textbf{100.0}   & \textbf{92.5}  & \textbf{100.0} & 66.6 & 79.9 \\
      LinkedList      &  100.0  & \textbf{100.0}  & \textbf{100.0}  & 100.0 & 23.7  & 38.4 \\
      NFST            &  \textbf{87.2} & \textbf{89.2}   & \textbf{88.2}   & 54.1 & 70.2 & 61.1 \\
      Signature       &   100.0 & \textbf{100.0}  & \textbf{100.0}    & 100.0 & 91.2 & 95.4 \\
      Socket          &  \textbf{87.3}  & \textbf{67.3}  & \textbf{76.0}  & 58.4 & 62.6 & 60.4 \\
      StackAr         & \textbf{86.8}   & 93.9  & \textbf{89.8} &  61.6 & \textbf{97.1} & 75.4 \\
      StringTokenizer &  \textbf{100.0} & 100.0 & \textbf{100.0} &  93.6 & 100.0 & 96.7 \\
      ZipOutputStream & \textbf{100.0}  & 100.0 & \textbf{100.0}   & 80.6 & 100.0 & 89.3 \\
      \hline
      Average         &  \textbf{87.8}  & \textbf{88.3}  & \textbf{87.8} &  77.3 & 67.3 & 68.4 \\
      \hline

  \end{tabular}
\end{table}

We initialized DICE using the test suite used by DSM in its evaluation, 
which is generated by Randoop~\cite{pacheco2007feedback}. 
From the results reported in Table \ref{tab:fsm_dsm_max}, 
DICE improves on the average F-measure of DSM by over 19\% (from 68.4 to 87.8),
and, for every class, the difference between DICE and DSM is statistically significant -- measured using the Wilcoxon signed-rank test.
This indicates that DICE was effective in inferring FSA models.

\begin{table}
  \caption{Precision, Recall, and F-measure of Tautoko and DICE. NFST refers to NumberFormatStringTokenizer.}
    \label{tab:tautoko}
    \begin{tabular}{ |c | ccc | ccc|}
      \hline
      \multicolumn{1}{|c|}{\multirow{2}{*}{\textbf{Class}}}  &  \multicolumn{3}{c|}{\textbf{DICE}} & \multicolumn{3}{c|}{\textbf{Tautoko}} \\ 
\cline{2-7} 
\multicolumn{1}{|c|}{}                      
 & \textbf{P} & \textbf{R} & \textbf{F}  & \textbf{P} & \textbf{R} & \textbf{F}        \\ \hline
      ArrayList       &  31.4   & 27.3   &  29.2 &  - & - & - \\
      HashMap         &  \textbf{100.0}  & \textbf{94.1}   & \textbf{97.0} & 56.7 & 25.4 & 35.1  \\
      HashSet         &  \textbf{87.4}   & \textbf{100.0}  & \textbf{93.3} & 100.0 & 13.6 & 23.9 \\
      Hashtable       &  \textbf{84.0}   & \textbf{100.0}  & \textbf{92.5}  & 38.8 & 23.3 & 29.1 \\
      LinkedList      &  100.0  & \textbf{100.0}  & \textbf{100.0}  & 100.0 & 20.9 & 34.6 \\
      NFST            &  87.2   & 89.2   & 88.2   & \textbf{100.0} & \textbf{100.0} & \textbf{100.0} \\
      Signature       &  100.0  & \textbf{100.0}  & \textbf{100.0}    & 100.0 & 23.8 & 38.4 \\
      Socket          &  \textbf{87.3}   & \textbf{67.3}   & \textbf{76.0}  & 84.1 & 24.4 & 37.7 \\
      StackAr         &  86.8   & \textbf{93.9}   & 89.8 & \textbf{100.0} & 87.0 & \textbf{92.8} \\
      StringTokenizer &  100.0  & 100.0  & 100.0 & 100.0  & 100.0  &100.0 \\
      ZipOutputStream &  100.0  & \textbf{100.0}  & \textbf{100.0}   & 100.0 & 25.0 & 40.5 \\
      \hline
      Average         &  87.8  &88.3  & 87.8 & 88.0  & 44.3 & 53.2 \\
      \hline

  \end{tabular}
\end{table}

We also investigated the effectiveness of DICE against Tautoko~\cite{dallmeier2010generating}, which generates tests based on missing transitions in an initial model. 
We compare the FSAs mined by DICE against Tautoko's.
The publicly available version of the executable artifact was downloaded from Tautoko's website\footnote{https://www.st.cs.uni-saarland.de/models/tautoko/} 
and executed on the same evaluation classes above, using Randoop generated tests as input to Tautoko. 
For some classes, Tautoko produced models containing methods that were not present in the ground-truth models. 
We therefore manually modified the models produced by Tautoko such that these methods were omitted, and merged states 
connected by a transition labelled with removed methods.  
For Socket, ZipOutputStream, and Signature, we evaluate the models published on Tautoko's homepage 
due to technical difficulties we encountered trying to run Tautoko on these classes.
However, we modify the evaluation criteria as Tautoko does not produce models with transitions 
labelled with boolean return values of method calls.
To compute the evaluation metrics for Tautoko's models, we ignore return values. 
This should generally lead to higher F-measures. 
We report the results in Table \ref{tab:tautoko}.

In one case (ArrayList), Tautoko does not run to completion within 24 hours.
For the other classes that Tautoko can mine models for, we observe that Tautoko does not produce models of high F-measures. 
Apart from StackAr, DICE outperforms Tautoko on the 11 classes.
While DICE produces models with an average F-measure of 87.8, Tautoko produces models with an average F-measure of 53.2.
If we omit the model of ArrayList, then DICE produces models with an average F-measure of 93.7.
We hypothesize that in certain cases, Tautoko's reliance on inspector methods (see Section \ref{sec:state_abstraction}) meant that it can not identify 
the right abstract states. 
For example, ZipOutputStream's state is not characterized by any inspector methods, and as such, Tautoko is unable to 
mine a good model of it. 

\vspace{0.2cm}\noindent\fbox{%
    \parbox{\textwidth}{%
        The adversarial specification mining process implemented by DICE produces FSA-based models of higher quality, 
        which outperforms existing approaches for inferring FSAs
    }%
}

\subsection{RQ2}

To answer RQ2, we aim to determine if the improvements was a result of our improvements to Evosuite, 
by studying if Evosuite alone was enough to produce diverse tests that would benefit the specification miner.
We use Evosuite (version 1.0.6), with the DynaMOSA~\cite{panichella2017automated} search algorithm, as a baseline approach, 
collecting the traces produced by the final test suite that is the output of Evosuite.
We use the default configuration of Evosuite. 
We do not try to find the optimal configuration for Evosuite as previous studies have indicated that tuning 
these parameters often do not outperform the default configuration~\cite{arcuri2013parameter}.
By default, the population size of test cases is 50 individuals. 
The default crossover operator is used, which is the single-point crossover with probability of 0.75. 
The selection of test cases is done using tournament selection, with a tournament size of 10. 
Tests are mutated with a probability inversely proportional to the number of statements it contains. 
We use the same test budget of 15 minutes for both DICE-Tester and Evosuite.
The traces produced from Evosuite are passed as input to DICE-Miner instead of the traces from DICE-Tester, and 
we compute the evaluation metrics of the FSA models produced.

\begin{table}
  \caption{Precision, Recall, and F-measure of DICE-Miner when using DICE-Tester and Evosuite. NFST refers to NumberFormatStringTokenizer. * indicate that the difference in F-measures is statistically significant.}
    \label{tab:testing_comparison}
    \begin{tabular}{ |c | ccc | ccc|}
      \hline
      \multicolumn{1}{|c|}{\multirow{2}{*}{\textbf{Class}}}  &  \multicolumn{3}{c|}{\textbf{DICE-Tester}} & \multicolumn{3}{c|}{\textbf{Evosuite}} \\ 
\cline{2-7} 
\multicolumn{1}{|c|}{}                      
 & \textbf{P} & \textbf{R} & \textbf{F}  & \textbf{P} & \textbf{R} & \textbf{F}        \\ \hline
      ArrayList *      &  31.4  & \textbf{27.3}  &  \textbf{29.2} &  \textbf{74.7} & 17.0		& 27.6 \\
      HashMap         &  100.0  &   94.1& 97.0 &  100.0 & 94.1   & 97.0 \\
      HashSet *      &  \textbf{87.4}  & \textbf{100.0}   & \textbf{93.3} &  84.5 & 65.2 & 74.3  \\
      Hashtable       &  84.0  & \textbf{100.0}   & \textbf{92.5}  & \textbf{91.0} & 93.1 & 92.0  \\
      LinkedList *     &  100.0  & \textbf{100.0}  & \textbf{100.0}  &  100.0 & 89.9 & 94.7  \\
      NFST            &  87.2 & 89.2   & 88.2   &  87.2 & 89.2 & 88.2 \\
      Signature       &   100.0 & 100.0  & 100.0    &  100.0 & 100.0 & 100.0  \\
      Socket          &  \textbf{87.3}  & \textbf{67.3}  & \textbf{76.0}  & 86.4 & 67.5 & 75.8  \\
      StackAr *        & \textbf{86.8}   & \textbf{93.9}  & \textbf{89.8} &  68.9 & 84.7 & 76.0  \\
      StringTokenizer &  100.0 & 100.0 & 100.0 &   100.0  &  100.0 & 100.0  \\
      ZipOutputStream & 100.0  & 100.0 & 100.0   & 100.0 & 100.0 & 100.0 \\
      \hline
      Average         &  87.8  & \textbf{88.3}  & \textbf{87.8} &   \textbf{90.4}  & 81.9  & 84.1   \\
      \hline

  \end{tabular}
\end{table}

The results are shown in Table \ref{tab:testing_comparison}.
On some classes, DICE-Tester can outperform Evosuite by up to 18.9\% in F-measure,
and on average, DICE-Tester outperforms Evosuite by about 3.7\%. 
To mitigate the effect of randomness, we run the experiments 20 times and compute if the differences are statistically significant using the Wilcoxon signed-rank test. 
We find that the differences are statistically significant for 4 out of the 11 classes in the benchmark.
This indicates that, on its own, Evosuite is already effective in generating diverse test cases,
although DICE-Tester can explore some uncommon usage patterns more effectively. 
We hypothesize that many of these classes do not exhibit complex usage constraints, 
therefore, Evosuite already performs well for these classes. 
However, to effectively explore non-trivial usage constraints, DICE-Tester can help significantly.

\vspace{0.2cm}\noindent\fbox{%
    \parbox{\textwidth}{%
        While Evosuite, with the DynaMOSA algorithm, is already able to aid the specification mining process by producing traces that allow DICE-Miner 
        to achieve good performance, DICE-Tester takes it a step further, producing traces that are even better.
    }%
}

\subsection{RQ3}

\begin{table}
  \caption{Precision, Recall, and F-measure of DICE and DSM with the traces produced from DICE-Tester. NFST refers to NumberFormatStringTokenizer.}
    \label{tab:fsm_dsm_with_extra_traces}
    \begin{tabular}{ |c | ccc | ccc|}
      \hline
      \multicolumn{1}{|c|}{\multirow{2}{*}{\textbf{Class}}}  &  \multicolumn{3}{c|}{\textbf{DICE}} & \multicolumn{3}{c|}{\textbf{DSM  }} \\ 
\cline{2-7} 
\multicolumn{1}{|c|}{}                      
 & \textbf{P} & \textbf{R} & \textbf{F}  & \textbf{P} & \textbf{R} & \textbf{F}        \\ \hline
      ArrayList       &  31.4  & \textbf{27.3}  &  \textbf{29.2} &  \textbf{60.4} & 16.5		& 25.9 \\
      HashMap         &  \textbf{100.0}  &   \textbf{94.1} & \textbf{97.0} &  30.8 &  86.0 & 45.3\\
      HashSet         &  \textbf{87.4}  & \textbf{100.0}   & \textbf{93.3} &  50.9 & 52.7 & 51.8 \\
      Hashtable       &  84.0  & \textbf{100.0}   & \textbf{92.5}  & \textbf{93.3} & 70.2 & 80.1 \\
      LinkedList      &  100.0  & \textbf{100.0}  & \textbf{100.0}  & 100.0  & 16.5 & 25.9 \\
      NFST            &  \textbf{87.2} & \textbf{89.2}   & \textbf{88.2}   & 57.3 & 81.9 & 67.4 \\
      Signature       &   100.0 & 100.0  & 100.0    & 100.0  & 100.0 & 100.0 \\
      Socket          &  \textbf{87.3}  & \textbf{67.3}  & \textbf{76.0}  & 40.7 & 63.9 & 49.8 \\
      StackAr         & \textbf{86.8}   & 93.9  & \textbf{89.8} &  47.2 & \textbf{100.0} & 64.1 \\
      StringTokenizer &  \textbf{100.0} & 100.0 & \textbf{100.0} &  75.3   & 100.0 & 85.9 \\
      ZipOutputStream & \textbf{100.0}  & \textbf{100.0} & \textbf{100.0}   & 79.8 & 75.4 & 77.5 \\
      \hline
      Average         &  \textbf{87.8}  & \textbf{88.3}  & \textbf{87.8} &   64.7  & 71.4  & 63.4  \\
      \hline

  \end{tabular}
\end{table}

To answer RQ3, we compare DICE and DSM with both approaches using the same set of traces.
We run DSM when provided with the traces from DICE-Tester. 
We compare the performance of DICE-Miner against DSM.
In this study, we do not compare our tool against other specification miners since DSM~\cite{le2018deep} 
is the state-of-the-art specification miner and 
has been demonstrated to outperform 
multiple approaches such as the traditional k-tails algorithm, SEKT, TEMI and CONTRACTOR++~\cite{le2018deep}.  
The results of using DSM with the additional traces are shown in Table \ref{tab:fsm_dsm_with_extra_traces}.

We observe that DSM's performance does not improve with the additional traces. 
In fact, the additional traces causes the performance of DSM to decrease. 
As DSM's states are partially determined by the probability of observing each next transition, 
we hypothesize that it is sensitive to the set of input traces used 
and it does not handle low-probability, but still valid, transitions robustly. 
Also noteworthy is that DICE-Miner can achieve a 100\% F-measure in 4 of the 11 classes, while DSM achieves a perfect score in only one case.
This shows that DICE-Miner is able to utilize diverse traces more effectively than DSM. 

\vspace{0.2cm}\noindent\fbox{%
    \parbox{\textwidth}{%
        DICE-Miner outperforms a state-of-the-art specification mining technique even when  
        the diverse traces produced by DICE-Tester are included in its input.
    }%
}

\section{Discussion}
\label{sec:discussion}

To further investigate our results, 
we raised four additional research questions for investigation and qualitatively analysed the models produced by DICE.

\begin{itemize}
  \item \textbf{RQ4: How effective were our adaptations of Evosuite for finding counterexamples?} \\
  RQ4 studies the effectiveness of DICE-Tester in discovering counterexamples of temporal properties. 
  Instead of measuring the quality of the FSAs output from DICE-Miner, an indirect measurement of how much the test generation
  benefited the specification mining process,
  we directly inspect the number of false temporal properties that the two tools, DICE-Tester and Evosuite (with DynaMOSA), are able to invalidate by finding counterexamples.

  \item \textbf{RQ5: How much did the constraints we introduced help to improve the performance of our specification mining algorithm?} \\
  RQ5 investigates the two constraints that we added in DICE-Miner motivated by the two observations we made.
  These observations are described in Section \ref{sec:approach_first_step}, 
  involving method purity and incompatible transitions on a state.
  The first constraint ensures that transitions labelled with side-effect free methods are self-loops,
  and the second constraint prevents the erroneous conflation of states that may accept traces that violate previously-mined temporal properties.  
  We try to drop these constraints and observe their effect on the performance of DICE-Miner. 

  \item \textbf{RQ6: How much does the quality of the initial test suite affect DICE?} \\
  RQ6 varies the quality of the initial test suite by using reduced subsets of it as input to the DICE process.
  We aim to investigate if the quality of the initial test suite has an effect on the quality of the models inferred by DICE.

  \item \textbf{RQ7: Can the FSAs inferred by DICE be used to support additional testing activities, for example, to perform protocol fuzzing?} \\
  RQ7 investigates if the FSAs learned by DICE can be used to aid in fuzzing servers of stateful protocols. 
  Effective fuzzing of a server requires the fuzzer to be aware of the specific order of messages to reach certain states. 
  We use state models learned by DICE to initialize a server fuzzer and observe if it helps the fuzzing process.
  This functions as an evaluation of DICE to determine if its inferred models have practical applicability.

\end{itemize}

\subsection{RQ4}

To try to further quantify the difference in performance of DICE-Tester and Evosuite+DynaMOSA, 
we compare the set of temporal properties that were successfully invalidated at the end of the test generation process.
We evaluate them against the ground-truth properties annotated by human experts.
These ground truth properties were made publicly available by Sun et al.~\cite{sun2019mining}.
The human experts annotated each possible temporal property following the template indicating if the property is true. 
This was done for three classes, HashSet, StringTokenizer and StackAr, which we use in our evaluation. 

As input to DICE-Tester, we enumerate all possible temporal properties between the methods of the class, 
and run DICE-Tester. 
For Evosuite+DynaMOSA, during the test generation process, we collect the traces when executing tests 
and print to standard output if the trace of the test produced is a counterexample to a temporal property.
Note that the set of input temporal properties 
will include properties that contradict each other (e.g. both \texttt{NF(A, B)} and \texttt{AF(A, B)} are part of the input).
In total, there were 56 true properties for HashSet out of 1014 possible properties, 
35 true properties out of 384 possible properties for StringTokenizer and 42 true properties  out of 600 possible properties for StackAr.

\begin{table}
  \caption{Number of incorrect rules failed to be invalidated when using all possible LTL properties as input}
    \label{tab:reduction}
    \begin{tabular}{|c|cc|}
      \hline
      Class&DICE-Tester&Evosuite\\
      \hline 
      HashSet & \textbf{51} &  233\\
      StackAr &  \textbf{39} &  108\\
      StringTokenizer & \textbf{35} & 216 \\
      \hline
      Average& \textbf{41.7} &  185.7\\
      \hline

  \end{tabular}
\end{table}

The results are reported in Table \ref{tab:reduction}, and we observe that in each class, 
DICE-Tester successfully found counterexamples for a vast majority of the incorrect properties. 
Out of an average of 621.7 incorrect properties, DICE-Tester successfully constructs tests 
that contradict an average of 580 of them, or over 93\% of the them.
In contrast, Evosuite does not succeed in invalidating most of the incorrect properties, 
and there were four times the number of incorrect properties that Evosuite failed to find counterexamples of.

\begin{table}
  \caption{Precision, Recall, and F-measure for DICE varying the threshold for resetting the test population. NFST refers to NumberFormatStringTokenizer. }
    \label{tab:initial_reset_threshold}
    \begin{tabular}{ |c | c | ccc | ccc|}
      \hline
      \multicolumn{1}{|c}{\multirow{2}{*}{\textbf{Class}}}  & \multicolumn{1}{|c|}{\textbf{100}} & \multicolumn{3}{c|}{\textbf{50}} & \multicolumn{3}{c|}{\textbf{150}} \\ 
\cline{2-8} 
\multicolumn{1}{|c|}{}                      
 & \textbf{F}         & \textbf{P} & \textbf{R} & \textbf{F}  & \textbf{P} & \textbf{R} & \textbf{F}        \\ \hline
      ArrayList       & \textbf{29.2}   & 	72.6 &15.7   &25.8  & 78.0 &16.1		& 26.6 \\
      HashMap         & 97.0   &  100.0 & 94.1 & 97.0 & 100.0  & 100.0	& \textbf{100.0} \\
      HashSet         & \textbf{93.2}   &  77.9 & 18.3 & 30.0 & 95.9 & 80.7 & 87.6 \\
      Hashtable       & 92.5   &  98.2 & 100.0 & \textbf{99.1} & 96.6 & 100.0 & 98.3 \\
      LinkedList      & 100.0   &  5.9 & 44.4 & 10.4 & 100.0  & 100.0 & 100.0 \\
      NFST            & \textbf{88.2}  &  87.7& 86.4 & 87.0 & 86.6  & 85.6 & 86.1 \\
      Signature       & 100.0   & 100.0 & 100.0 & 100.0 & 100.0  & 100.0 & 100.0 \\
      Socket          & \textbf{76.0}   &  53.4& 66.0 & 59.0 & 80.5 & 66.0 & 72.5 \\
      StackAr         & \textbf{89.8}   & 41.7 & 88.2 & 56.6 &  83.2 & 84.7 & 84.0 \\
      StringTokenizer & 100.0   & 100.0 & 100.0 & 100.0 & 100.0   & 100.0 & 100.0 \\
      ZipOutputStream & \textbf{100.0}   & 28.6 & 100.0 & 44.4 & 72.7  & 100.0 & 84.2 \\
      \hline
      Average         & \textbf{87.8}   & 69.6  & 73.9 & 64.5 & 91.3  & 86.7  & 86.7 \\
      \hline

  \end{tabular}
\end{table}

Next, we also investigate the effect of the threshold for resetting the test population, described in Section 3.3.
On average, the test budget of 15 minutes allows to search to run for 505 generations. 
With a threshold of 100 generations, we observe an average of 1 reset for each run.
We varied the threshold and run DICE on the classes in the benchmark. 
The evaluation metrics are reported in Table \ref{tab:initial_reset_threshold}.
Our findings are that having too low a threshold adversely affects the quality of the models learned by DICE.
Decreasing the threshold from 100 generations to 50 generation caused a large drop in F-measure from 87.8\% to 64.4\%, over a 20\% decline.
On the other hand, increasing the threshold to 150 generations  caused a slight decrease in quality, from 87.8\% to 86.7\%.
The results suggest that, at least for this benchmark set of classes, the default value that we choose (100 generations) is reasonably good.

\vspace{0.2cm}\noindent\fbox{%
    \parbox{\textwidth}{%
        DICE-Tester outperforms Evosuite in finding counterexamples for incorrect temporal properties, 
        validating the benefits of our modifications for improving the search process.
    }%
}

\subsection{RQ5}

\begin{table}
  \caption{F-measure of DICE-Miner, without the constraints we identified. NFST refers to NumberFormatStringTokenizer. }
    \label{tab:fsm_ablation}
    \begin{tabular}{|c | cccc|}
      \hline
      \textbf{Class} & \textbf{Both} & \textbf{Constraint A, w/o B} & \textbf{Constraint B, w/o A} & \textbf{None} \\
      \hline
      ArrayList       & 29.2   & 28.6     & 30.4  & 31.3 \\
      HashMap         & 97.0   &  97.0    & 97.8  & 97.8 \\
      HashSet         & 93.2   &  88.5    & 91.8  & 92.3 \\
      Hashtable       & 92.5   &  86.3    & 84.9  & 87.7 \\
      LinkedList      & 100.0  &  100.0   & 70.8  & 70.8 \\
      NFST            & 88.2   & 72.7     & 81.8  & 81.8 \\
      Signature       & 100.0  & 100.0    & 100.0 & 100.0 \\
      Socket          & 76.0   &  63.5    & 67.9  & 67.8  \\
      StackAr         & 89.8   &  71.9    & 65.2  & 74.5 \\
      StringTokenizer & 100.0 & 100.0    & 100.0 & 100.0 \\
      ZipOutputStream & 100.0   & 100.0     & 100.0  &  100.0 \\
      \hline
      Average         & 87.8   & 82.6     & 80.9  & 82.2 \\
      \hline

  \end{tabular}
\end{table}

To study if the two constraints described in Section \ref{sec:approach_first_step} influenced the performance of DICE-Miner, 
we ran more experiments, omitting the constraints.
Constraint A refers to the constraint that pure methods do not cause a transition to another state, while 
Constraint B refers to the constraint that a state should be split up if it has incompatible transitions.
We use the execution traces from Le et al.~\cite{le2018deep} and DICE-Tester in these experiments.

The results are presented in Table \ref{tab:fsm_ablation}. 
Without both constraints, the average F-measure dropped by about 5.2\%. 
In virtually all classes, we see a  decline in the performance of DSM-Miner.
We see that using information about method purity is important to achieving good performance. 
While using Constraint B alone did not provide any improvements without Constraint A, 
it was necessary to increase the performance to 87.8 from 82.6 (with only Constraint A).
This confirms our observations that these two constraints are important for specification mining.

\vspace{0.2cm}\noindent\fbox{%
    \parbox{\textwidth}{%
        The two constraints that we added for addressing the two observations were helpful for DICE-Miner to achieve its performance.        
    }%
}

\subsection{RQ6}

\begin{table}
  \caption{Precision, Recall, and F-measure for DICE varying the initial test suite. NFST refers to NumberFormatStringTokenizer. }
    \label{tab:initial_test_suite_ablation}
    \begin{tabular}{ |c | c | ccc | ccc|}
      \hline
      \multicolumn{1}{|c}{\multirow{2}{*}{\textbf{Class}}}  & \multicolumn{1}{|c|}{\textbf{100\%}} & \multicolumn{3}{c|}{\textbf{50\%}} & \multicolumn{3}{c|}{\textbf{25\%}} \\ 
\cline{2-8} 
\multicolumn{1}{|c|}{}                      
 & \textbf{F}         & \textbf{P} & \textbf{R} & \textbf{F}  & \textbf{P} & \textbf{R} & \textbf{F}        \\ \hline
      ArrayList       & \textbf{29.2}   &  31.9	& 21.1  & 25.4 & 28.6 &	25.3	& 26.8 \\
      HashMap         & 97.0   &  100.0	  & 94.1	& 97.0 & 100.0  &	83.1	& 90.8 \\
      HashSet         & 93.2   &  92.4  & 100.0   & 96.0 &94.1  & 100.0   & \textbf{97.0} \\
      Hashtable       & 92.5   &  98.0  & 100.0   & 99.0 & 96.3 & 100.0   & \textbf{98.1} \\
      LinkedList      & 100.0   &  100.0   & 100.0   & 100.0  & 100.0  & 100.0   & 100.0  \\
      NFST            & 88.2  &  92.1  &  89.4 & 90.7 & 92.9 & 87.9  & \textbf{90.3} \\
      Signature       & 100.0   &  100.0   & 100.0   & 100.0  &100.0   & 100.0   & 100.0  \\
      Socket          & \textbf{76.0}   &  84.8  & 62.5  & 72.0 & 88.3 & 61.7  & 72.6 \\
      StackAr         & \textbf{89.8}   &  76.4  & 81.9  & 79.0 & 73.9 & 84.7  & 78.9 \\
      StringTokenizer & 100.0   &  100.0   & 100.0   & 100.0  &100.0   &100.0    & 100.0  \\
      ZipOutputStream & 100.0   &  100.0   & 100.0   & 100.0  &100.0   &100.0    & 100.0  \\
      \hline
      Average         & \textbf{87.8}   &  88.7  & 86.3   & 87.2  & 88.6 & 85.7 & 86.8 \\
      \hline

  \end{tabular}
\end{table}

To answer RQ6, we investigate how sensitive the DICE process is to the initial input test suite. 
We performed experiments using different subsets of the initial test suite. 
We do not run experiments using the test suite originally written by the developers accompanying the classes in the benchmark. 
We manually analysed the test cases for these classes and found that exercising these tests would only produce a few traces. 
The functionality of each class is well tested, but the tests typically do not have a high diversity regarding the sequences of method invocations. 
For example, the test cases for a LinkedList exercise each method of the LinkedList, but do not show how the methods relate to one another. 
Therefore, we do not expect the evaluation metrics of DICE to change significantly from using a small subset (e.g. 25\%) of the initial test suite used as input to the earlier experiments.

The results are shown in Table \ref{tab:initial_test_suite_ablation}. 
By using 50\% of the initial test suite, F-measure drops from 87.8\% to 87.2\%, and by using 25\% of the initial test suite, it drops further to 86.8\%. 
This indicates that the initial quality of the test suite influences the quality of the models inferred by DICE, however, the difference 
is small.
Interestingly, we noticed that F-measure increased for some of the classes in the benchmark. 
The models for ArrayList and HashSet improved in quality by about 1\% in F-measure.
For these cases, 
as having fewer initial traces may mean that we infer a larger number of incorrect temporal properties, 
we hypothesize that these large number of incorrect traces can sometimes lead the search process to collaterally cover more temporal properties 
and produce informative traces that were useful to the inference process.
We answer RQ6 by concluding that the quality of the initial test suite has an impact on the models inferred by DICE, but overall, this impact is small.

\vspace{0.2cm}\noindent\fbox{%
    \parbox{\textwidth}{%
        The quality of initial test suite has a small effect (1\% change in the average F-measure) on the quality of the models inferred by DICE.
    }%
}

\subsection{Qualitative Evaluation}

While the DICE system results in an improved F-measure compared to existing approaches, it is not able to achieve 
100\% correct finite-state automata for all of the 11 classes. 
We manually inspected the resulting FSA to try to identify reasons as to why this is the case, and to propose next steps for our work.

\begin{figure}
  \begin{center}
    
      \begin{tikzpicture}[>=latex',line join=bevel,scale=0.75]
        \pgfsetlinewidth{1bp}
      \pgfsetcolor{black}
        \draw [->] (86.65bp,62.848bp) .. controls (119.04bp,62.848bp) and (166.4bp,62.848bp)  .. (207.15bp,62.848bp);
        \definecolor{strokecol}{rgb}{0.0,0.0,0.0};
        \pgfsetstrokecolor{strokecol}
        \draw (154.94bp,70.348bp) node {StringTokenizer};
        \draw [->] (420.35bp,26.513bp) .. controls (434.62bp,32.811bp) and (455.19bp,41.885bp)  .. (481.72bp,53.594bp);
        \draw (450.84bp,52.348bp) node {END};
        \draw [->] (394.9bp,35.905bp) .. controls (393.11bp,46.101bp) and (395.8bp,55.697bp)  .. (402.99bp,55.697bp) .. controls (407.59bp,55.697bp) and (410.35bp,51.758bp)  .. (411.07bp,35.905bp);
        \draw (402.99bp,63.197bp) node {hasMore:F};
        \draw [->] (244.53bp,67.967bp) .. controls (250.43bp,69.443bp) and (257.02bp,70.906bp)  .. (263.14bp,71.848bp) .. controls (332.94bp,82.596bp) and (351.59bp,87.153bp)  .. (421.84bp,79.848bp) .. controls (438.02bp,78.165bp) and (455.74bp,74.634bp)  .. (480.58bp,68.791bp);
        \draw (402.99bp,90.348bp) node {END};
        \draw [->] (244.87bp,58.223bp) .. controls (275.84bp,50.511bp) and (338.2bp,34.981bp)  .. (384.66bp,23.412bp);
        \draw (314.64bp,60.348bp) node {hasMore:F};
        \draw [->] (220.04bp,80.668bp) .. controls (218.88bp,90.575bp) and (220.96bp,99.697bp)  .. (226.29bp,99.697bp) .. controls (229.62bp,99.697bp) and (231.69bp,96.133bp)  .. (232.54bp,80.668bp);
        \draw (226.29bp,107.2bp) node {hasMore:T};
        \draw [->] (216.45bp,79.131bp) .. controls (209.89bp,97.129bp) and (213.17bp,117.7bp)  .. (226.29bp,117.7bp) .. controls (237.06bp,117.7bp) and (241.2bp,103.86bp)  .. (236.14bp,79.131bp);
        \draw (226.29bp,125.2bp) node {nextToken};
      \begin{scope}
        \definecolor{strokecol}{rgb}{0.0,0.0,0.0};
        \pgfsetstrokecolor{strokecol}
        \draw (102.67bp,74.4bp) -- (63.37bp,74.4bp) -- (51.22bp,111.78bp) -- (39.08bp,74.4bp) -- (-0.22bp,74.4bp) -- (31.57bp,51.3bp) -- (19.43bp,13.92bp) -- (51.22bp,37.02bp) -- (83.02bp,13.92bp) -- (70.87bp,51.3bp) -- cycle;
        \draw (51.222bp,62.848bp) node {C4};
      \end{scope}
      \begin{scope}
        \definecolor{strokecol}{rgb}{0.0,0.0,0.0};
        \pgfsetstrokecolor{strokecol}
        \draw (226.29bp,62.85bp) ellipse (18.7bp and 18.7bp);
        \draw (226.29bp,62.848bp) node {C0};
      \end{scope}
      \begin{scope}
        \definecolor{strokecol}{rgb}{0.0,0.0,0.0};
        \pgfsetstrokecolor{strokecol}
        \draw (502.69bp,62.85bp) ellipse (18.72bp and 18.72bp);
        \draw (502.69bp,62.85bp) ellipse (22.7bp and 22.7bp);
        \draw (502.69bp,62.848bp) node {C3};
      \end{scope}
      \begin{scope}
        \definecolor{strokecol}{rgb}{0.0,0.0,0.0};
        \pgfsetstrokecolor{strokecol}
        \draw (402.99bp,18.85bp) ellipse (18.7bp and 18.7bp);
        \draw (402.99bp,18.848bp) node {C2};
      \end{scope}
      \end{tikzpicture}
    
  \end{center}
  \caption{Example of a FSA model produced by DSM. hasMore:T is short for hasMoreTokens:TRUE and hasMore:F is short of hasMoreTokens:FALSE}
  \label{fig:100_pc_not_eq}
\end{figure}
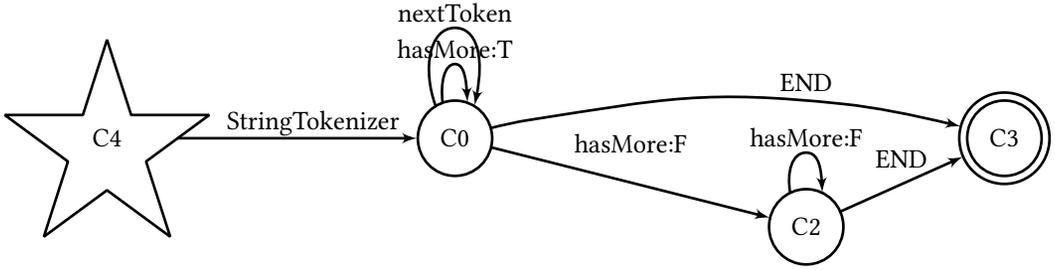

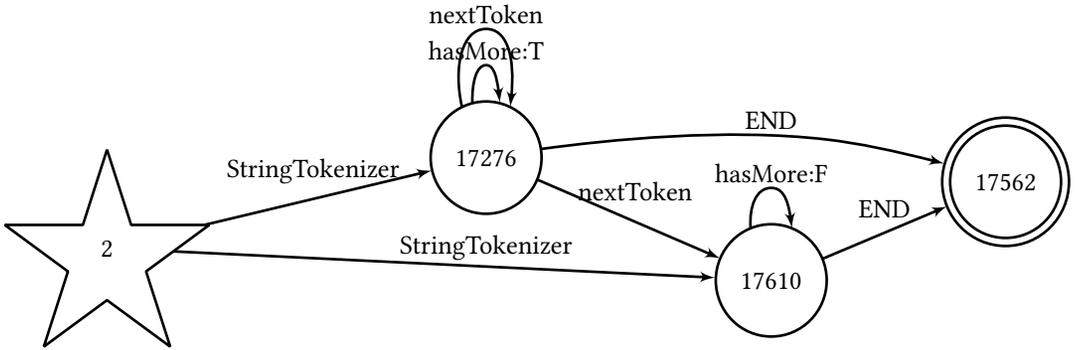
\begin{figure}
  \begin{center}
    
\begin{tikzpicture}[>=latex',line join=bevel,scale=0.75]
  \pgfsetlinewidth{1bp}
\pgfsetcolor{black}
  \draw [->] (100.34bp,60.303bp) .. controls (132.82bp,67.967bp) and (174.75bp,77.86bp)  .. (214.52bp,87.245bp);
  \definecolor{strokecol}{rgb}{0.0,0.0,0.0};
  \pgfsetstrokecolor{strokecol}
  \draw (154.94bp,87.215bp) node {StringTokenizer};
  \draw [->] (84.485bp,47.122bp) .. controls (146.85bp,44.136bp) and (281.01bp,37.713bp)  .. (357.05bp,34.072bp);
  \draw (241.94bp,49.215bp) node {StringTokenizer};
  \draw [->] (269.64bp,98.001bp) .. controls (303.6bp,102.61bp) and (362.87bp,108.53bp)  .. (413.34bp,102.71bp) .. controls (429.67bp,100.83bp) and (447.35bp,97.09bp)  .. (472.37bp,90.782bp);
  \draw (385.39bp,112.21bp) node {END};
  \draw [->] (235.02bp,121.07bp) .. controls (234.79bp,131.27bp) and (237.1bp,139.66bp)  .. (241.94bp,139.66bp) .. controls (245.05bp,139.66bp) and (247.11bp,136.22bp)  .. (248.87bp,121.07bp);
  \draw (241.94bp,147.16bp) node {hasMore:T};
  \draw [->] (229.78bp,118.99bp) .. controls (225.39bp,138.19bp) and (229.44bp,157.66bp)  .. (241.94bp,157.66bp) .. controls (252.2bp,157.66bp) and (256.77bp,144.56bp)  .. (254.1bp,118.99bp);
  \draw (241.94bp,165.16bp) node {nextToken};
  \draw [->] (267.92bp,82.668bp) .. controls (290.97bp,72.865bp) and (324.85bp,58.461bp)  .. (359.61bp,43.679bp);
  \draw (316.94bp,77.215bp) node {nextToken};
  \draw [->] (411.26bp,43.468bp) .. controls (426.82bp,49.933bp) and (446.92bp,58.287bp)  .. (473.74bp,69.433bp);
  \draw (442.34bp,68.215bp) node {END};
  \draw [->] (374.98bp,58.694bp) .. controls (374.27bp,69.501bp) and (377.73bp,78.662bp)  .. (385.39bp,78.662bp) .. controls (390.42bp,78.662bp) and (393.64bp,74.717bp)  .. (395.8bp,58.694bp);
  \draw (385.39bp,86.162bp) node {hasMore:F};
\begin{scope}
  \definecolor{strokecol}{rgb}{0.0,0.0,0.0};
  \pgfsetstrokecolor{strokecol}
  \draw (102.67bp,60.27bp) -- (63.37bp,60.27bp) -- (51.22bp,97.65bp) -- (39.08bp,60.27bp) -- (-0.22bp,60.27bp) -- (31.57bp,37.16bp) -- (19.43bp,-0.22bp) -- (51.22bp,22.89bp) -- (83.02bp,-0.22bp) -- (70.87bp,37.16bp) -- cycle;
  \draw (51.222bp,48.715bp) node {2};
\end{scope}
\begin{scope}
  \definecolor{strokecol}{rgb}{0.0,0.0,0.0};
  \pgfsetstrokecolor{strokecol}
  \draw (241.94bp,93.71bp) ellipse (27.9bp and 27.9bp);
  \draw (241.94bp,93.715bp) node {17276};
\end{scope}
\begin{scope}
  \definecolor{strokecol}{rgb}{0.0,0.0,0.0};
  \pgfsetstrokecolor{strokecol}
  \draw (385.39bp,32.71bp) ellipse (27.9bp and 27.9bp);
  \draw (385.39bp,32.715bp) node {17610};
\end{scope}
\begin{scope}
  \definecolor{strokecol}{rgb}{0.0,0.0,0.0};
  \pgfsetstrokecolor{strokecol}
  \draw (503.29bp,81.71bp) ellipse (27.9bp and 27.9bp);
  \draw (503.29bp,81.71bp) ellipse (31.9bp and 31.9bp);
  \draw (503.29bp,81.715bp) node {17562};
\end{scope}
\end{tikzpicture}
\end{center}
\caption{Example of a FSA model produced by DICE-Miner. hasMore:T is short for hasMoreTokens:TRUE and hasMore:F is short for hasMoreTokens:FALSE}
\label{fig:100_pc_not_eq_1}
\end{figure}

One interesting observation is that the interpretation of FSA models may differ between models that have identical F-measures.
Apart from boosting the accuracy of models, 
having knowledge of pure methods will produce models that are qualitatively better for program comprehension.
For example, referring to Figure \ref{fig:100_pc_not_eq}, DSM produces a model for StringTokenizer that suggests that \texttt{nextToken} is always enabled should a developer never invoke \texttt{hasMoreTokens}. 
In contrast, the model produced by DICE-Miner, as shown in Figure \ref{fig:100_pc_not_eq_1}, cannot be erroneously interpreted in this manner. 
The invocation of \texttt{hasMoreTokens} with different return values are not allowed on the same state 
and it is clear that \texttt{nextToken} may change the state of a StringTokenizer object such that 
\texttt{nextToken} can not be further invoked. 
Note that both the models produced by DSM and DICE-Miner have an F-measure of 100, showing that there may be qualitative 
differences between models that are automatically evaluated to be perfectly accurate. 

The method \texttt{hasMoreTokens} represents an interesting case that may be a next step for DICE-Miner.
While it is accurately identified as a side-effect free method heuristically by DICE-Miner,
the static analysis we performed did not reveal it as a pure method.
The implementation of \texttt{hasMoreTokens} is, in fact, impure.
A static analysis-based approach therefore treats hasMoreTokens as an impure method.
More sophisticated analysis of purity, such as using the notion of observational-purity~\cite{barnett200499,naumann2007observational}  
will help to produce qualitatively better models.  
Observationally-pure methods refer to the class of methods, in which \texttt{hasMoreTokens} belongs to, 
which have side-effects that cannot be observed outside of the class. 
From the perspective of learning usage models and specifications about usage of the class, 
these methods are effectively pure. 
While our naive name-based heuristic may help to identify some of these methods, 
it is likely that there are observationally-pure methods in the wild that we cannot detect.

Investigating the poor performance of DICE for some classes, 
the lack of expressiveness of the six LTL property templates considered is a possible reason for it.  
LTL properties templates involving more than 2 events and the use of other temporal operators beyond the basic operators may be required.
For example, clients of NumberFormatStringTokenizer are able to reset its state using the \texttt{reset} method, 
as such, there are constraints between its methods that can only be expressed through the use of LTL formula involving more than 2 events.  
The property \texttt{NF(hasNextToken():FALSE,nextToken)} is false, contrary to intuition, since invoking \texttt{reset} after \texttt{hasNextToken:FALSE} 
may allow \texttt{nextToken} to be invoked again.
In this case, the three-event property \texttt{(hasNextToken():FALSE NF nextToken) U reset} is necessary to accurately represent the temporal constraints between the methods. 
This formula indicates that \texttt{nextToken} cannot follow \texttt{hasNextToken():FALSE} until the object instance has been \texttt{reset}. 

While in this work, we have considered only 2 event LTL formulae, 
it may be possible to use formulae relating 3 or more events and use them during the testing process 
or to guide the merging of the states in the automaton. 
However, having more complex formulae will come at a cost and there may be a trade-off; 
gaining some accuracy but slowing down the approach. 
Including longer rules will lead to an exponential  growth in their numbers. 
Indeed, many studies have focused on mining rules and patterns involving only 2 events~\cite{le2015beyond,yang2006perracotta,safyallah2006dynamic,lo2008mining}, 
and researchers have noted the problem of scalability when mining longer patterns~\cite{pei2004mining}.

\subsection{RQ7}

\subsubsection{Using the inferred FSAs for fuzzing}

Next, we evaluate models produced by DICE in its applicability on fuzzing servers of stateful network protocols~\cite{pham2020aflnet}. 
In fuzzing, random test inputs are generated automatically in order to find bugs. 
It is important to discover critical bugs in the implementation of protocols. 
The Heartbleed vulnerability~\footnote{\url{https://heartbleed.com/}}, 
a security bug in the implementation of the Transport Layer Security in the OpenSSL library,
has shown the pervasiveness and the high cost of such bugs~\cite{durumeric2014matter}. 
Server fuzzing may help in finding these bugs and several server fuzzers, such as AFLNET~\cite{pham2020aflnet}, have been proposed. 
Servers are stateful and reaching specific states may require specific sequences of messages between server and client. 
Without information about the specific messages required, the fuzzer is unlikely to send a sequence of messages that exercises program states deep in the server. 
The FSA model learned by DICE can be used as a state model to guide the server in reaching states that are difficult to reach otherwise.

AFLNET~\cite{pham2020aflnet}  is a coverage-guided fuzzer that has been shown to outperform other server fuzzers. 
Like other coverage-guided fuzzers, AFLNET instruments the program to receive feedback if there is an increase in the server’s code coverage achieved by an input. 
This feedback is used by the fuzzer to decide which inputs to mutate, optimising its choice of inputs for increased coverage of the program. 
Unlike other coverage-guided fuzzers, AFLNET has a state model of the server and detects if inputs lead to unexplored states. 
In other words, an input is retained for further mutation by AFLNET if it leads to increased coverage or enters an unexplored state. 
During the fuzzing process, AFLNET constructs a state model of the server using the observed status code sent by the server. 
By updating the state model during runtime, AFLNET is able to detect bugs in states that only appear in the implementation of the protocol 
and not in the protocol’s official definition. As a tradeoff, AFLNET may spend a significant amount of time early in the fuzzing process 
using a rudimentary state model and, until the state model is refined, is unable to reach deep into the program.

During the fuzzing process, AFLNET selects the next input for fuzzing by selecting a state in the state model to act as the target state.
From this target state, two criteria are applied.
First, AFLNET picks, with a higher probability, states that are less frequently exercised.
Second, states that have been associated with a greater number of inputs that 
lead to higher coverage or new states are more likely to be selected.
Then, an input associated with that state is mutated in a way that ensures that that particular state is still visited by the new input.

A state inference algorithm, such as DICE, can enhance the workings of AFLNET by providing it with the state model before starting the fuzzer. 
By running DICE on a protocol client, DICE produces a model of the protocol which can be used for server fuzzing. 
We observe that the API members of a protocol's client tend to have a one-to-one correspondence to the request types defined in a protocol.
For instance, an FTP client typically has API members for each request type (e.g. the \texttt{user()} method sends a USER request, the \texttt{pass()} method sends a PASS request).
Consequently, \textbf{an automaton specification of a protocol's client API is a model of the protocol from the client's perspective}. 
The state model is, therefore, useful for a fuzzer to simulate the protocol's client.
To target a particular state, one can traverse the edges from the starting state until the target state is reached; 
the labels on the edges traversed are the requests the client will have to make.

We modify the state-of-the-art server fuzzer, AFLNET, to use the automaton produced from DICE as the state model.
Instead of starting the fuzzing process with an empty state model, we modify it to be initialized with the output of DICE.
We also make some modifications in AFLNET related to how it uses a state model. 
Presently, the status code in the server's response is used as an indicator of the present state.
On the other hand, 
the states in DICE's state machine are more granular.
We modify AFLNET to traverse the state machine while considering more information than the response code of the server,
namely, the sequence of requests made so far.
Additionally, when the server responds with a status code indicating an error, 
we assume that the state has not changed. 
In contrast, AFLNET transits to an error state when the server replies with server code indicating an error. 
Overall, with our modifications and initialization with DICE's output, 
AFLNET can work with a higher granularity of states and is more likely to generate inputs that are accepted, 
allowing it to generate inputs that go deeper into the program. 
To distinguish between the two systems, we will refer to our modified version of AFLNET as DICE+AFLNET.

\begin{figure}[t]
	\centering
	\includegraphics[width=0.6\linewidth]{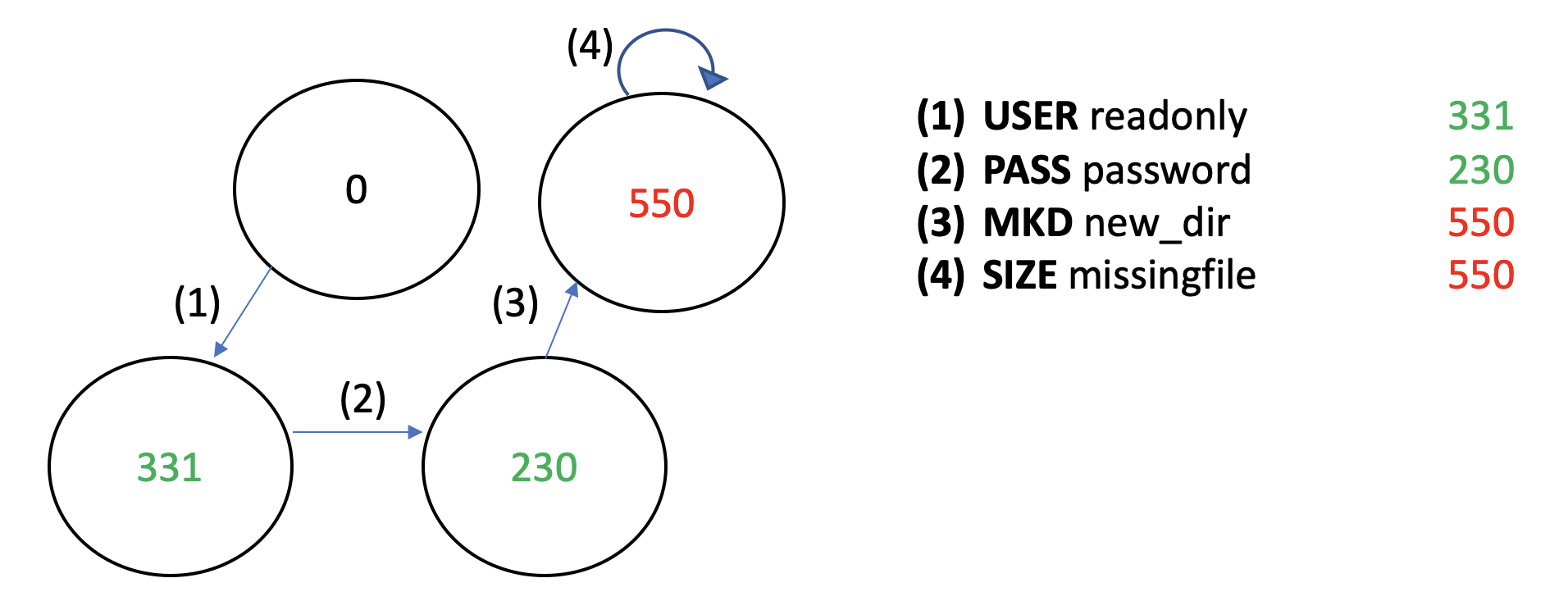}
	\caption{State traversal in AFLNET's inferred state machine for the given sequence of request types}
	\label{fig:aflnet_states}
\end{figure}

\begin{figure}[t]
	\centering
	\includegraphics[width=0.6\linewidth]{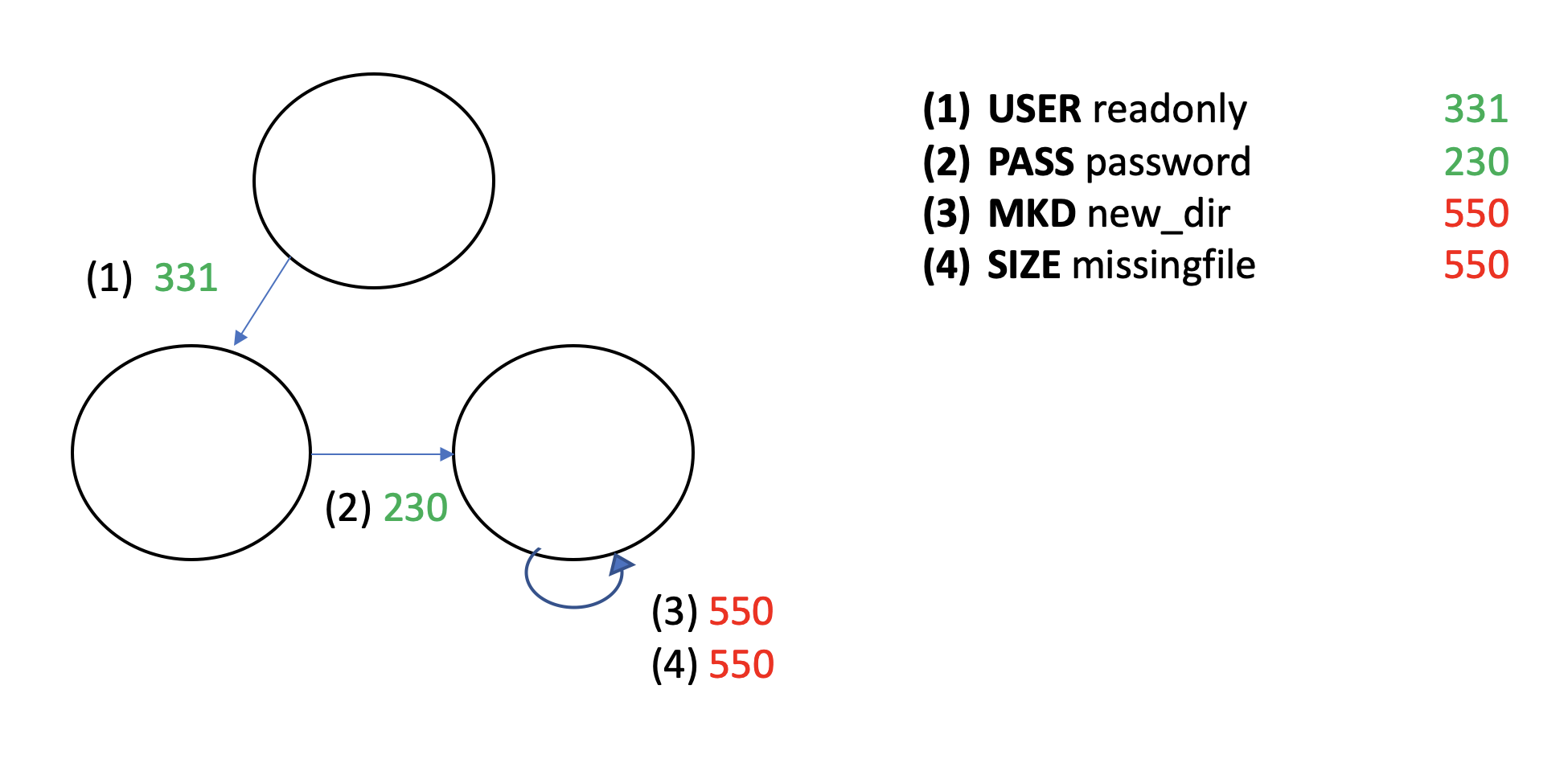}
	\caption{State traversal in DICE's inferred state machine for the same sequence of request types as Figure \ref{fig:aflnet_states}}
	\label{fig:dice_aflnet_states}
\end{figure}

To look at an example of the difference between AFLNET and DICE+AFLNET, 
we use the following example sequence of requests made by AFLNET: 
\texttt{[USER, PASS, MKD, SIZE]}.
In this sequence, the client has successfully logged in with a user.
The user then fails to create a new directory, due to insufficient access rights, 
and attempts to get the size of file that does not exist on the server.
Given this sequence of requests, the server responds with the following sequence of response codes:
\texttt{(331, 230, 550, 550)}, 
indicating that the login was successful but the creation of the new directory and access of a non-existent file have failed.
AFLNET traverses the states as shown in Figure \ref{fig:aflnet_states}, in which it first moves to state 331, then to 230, and to 550. 
On the other hand, DICE+AFLNET traverses the states as shown in Figure \ref{fig:dice_aflnet_states}. 
Notice that DICE+AFLNET remains in the same state after the failed requests.
Compared to the transition to state 550 used in AFLNET, we suggest that this more accurately captures the semantics of a failed request in the two protocols that we study.
As this may not be true of many network protocols,  
we will provide configuration options allowing the user of the fuzzer to specify the semantics of a failed request with respect to the state model.
Overall, this enables AFLNET to work with the state models from DICE, which are more granular.

\subsubsection{Experimental Results}

To determine if the models learned by DICE can aid server fuzzers, 
we evaluate DICE+AFLNET on two protocols that were studied in the evaluation of AFLNET~\cite{pham2020aflnet}, FTP and RTSP. 
We reuse the same FTP\footnote{\url{https://github.com/hfiref0x/LightFTP}} and RTSP\footnote{\url{http://www.live555.com/mediaServer/}} servers that were fuzzed by Pham et al.~\cite{pham2020aflnet}.
We run the modified version of AFLNet that takes a state machine from DICE as input.
For each of the two protocols, we performed a search on GitHub to find open-source clients\footnote{\url{https://github.com/apache/commons-net}}\footnote{\url{https://github.com/mutaphore/RTSP-Client-Server}} of the protocol.
We selected one client for each protocol and ran DICE on it.
This process is semi-automatic. 
Some human effort was required to annotate which API methods directly represent a request from the client.  
We modified the source of the clients to enable DICE to run effectively on it. 
Originally, the clients have different API usage patterns in which the developers have to check the integer return value of the method call
(\texttt{int returnCode = ftpClient.user(username)}), 
or invoke another method to retrieve the response server code (\texttt{if (parseServerResponse() != 200)}). 
We modified the clients to throw exceptions on requests where the server responds with status codes signalling failure. 
For each of the two clients, the modification of the client took less than 15 minutes. 
This was done for DICE to detect a failing method call. 
DICE was run on each client for another 15 minutes.

The modification of the client and annotation of the methods took less than 15 minutes for each protocol.
In total, less than 30 minutes is required to make the necessary modifications and to run DICE.
For a fair comparison, we grant AFLNET an additional hour of fuzzing to account for the additional effort and time to run DICE.
We fuzz each protocol for 24 hours using DICE+AFLNET, and 25 hours using the original version of AFLNET. 
We compare the fuzzers by the average line and branch coverage of the resulting inputs.
Apart from the additional step of running the automata inference algorithms and the different running time, 
we follow the same fuzzing procedure as prior work.
As the fuzzing process is stochatic, we mitigate the effect or randomness by running 20 independent experiments for each fuzzer and report the average coverage obtained.

\begin{table}
  \caption{Coverage achieved on each protocol server by the AFLNET and DICE+AFLNET. Numbers in parenthesis indicate the proportion of total lines/branches covered.}
    \label{tab:coverage}
    \begin{tabular}{| c | c | c | c | c| c | c |}
      \hline
      
\multicolumn{1}{|c|}{\multirow{2}{*}{\textbf{Protocol}}}  & \multicolumn{2}{c|}{\textbf{AFLNET}} & \multicolumn{2}{c|}{\textbf{DICE+AFLNET}} \\ 
\cline{2-5} 
\multicolumn{1}{|c|}{}                     & 
\textbf{Lines Covered}           & \textbf{Branches Covered}          & \textbf{Lines Covered}         & \textbf{Branches Covered}   \\ \hline
      FTP       & 644 (57\%) &  311 (40\%)    &  \textbf{777  (69\%)} & \textbf{400 (50\%)} \\
      RTSP      & 2453 (11\%)  &  1216 (7\%)    &  \textbf{2470 (11\%)} & \textbf{1234 (7\%)} \\
      \hline

  \end{tabular}
\end{table}

The average coverage achieved by the fuzzers  are shown in Table \ref{tab:coverage}. 
On the RTSP server, there was only a slight increase in coverage (of 17 lines and 18 branches).
As Pham et al.~\cite{pham2020aflnet} observed, the RTSP server has fewer states to explore and the number of messages required 
to exercise code deeper into the server is lower than FTP.
In other words, for such protocols where the state model is simpler, AFLNET benefits less from having a state model initialized before it begins fuzzing.
On the other hand, on the FTP server, 
the use of the state machine from DICE significantly improves both line and branch coverage.
The average line coverage increased from 57\% to 69\% and branch coverage increased from 40\% to 50\%. 

\begin{table}
  \caption{Coverage achieved on each protocol server when using the state models from DSM and DICE. Numbers in parenthesis indicate the proportion of total lines/branches covered.}
    \label{tab:coverage_vs_dsm}
    \begin{tabular}{| c | c | c | c | c| c | c |}
      
      \hline
\multicolumn{1}{|c|}{\multirow{2}{*}{\textbf{Protocol}}}  & \multicolumn{2}{c|}{\textbf{DSM+AFLNET}} & \multicolumn{2}{c|}{\textbf{DICE+AFLNET}} \\ 
\cline{2-5} 
\multicolumn{1}{|c|}{}                     & 
\textbf{Lines Covered}           & \textbf{Branches Covered}          & \textbf{Lines Covered}         & \textbf{Branches Covered}   \\ \hline
      FTP       & 703 (62\%) &  341 (43\%)    &  \textbf{777  (69\%)} & \textbf{400 (50\%)} \\
      RTSP      & 2448 (11\%)  &  1201 (7\%)    &  \textbf{2470 (11\%)} & \textbf{1234 (7\%)} \\
      \hline

  \end{tabular}
\end{table}

Next, we investigate if AFLNET achieves the same increase in performance when we use the automata inferred by DSM~\cite{le2018deep} instead of DICE. 
The experimental results are shown in Table \ref{tab:coverage_vs_dsm}.
On the RTSP server, the difference between the use of models from DSM and DICE is insignificant.
However, on the FTP server, DICE achieves a significantly higher line and branch coverage.
We also observe that the coverage on the FTP server increased over the baseline AFLNET fuzzer;
it increases from 57\% to 62\% line coverage,
suggesting that the modifications we made to initialize the fuzzing with an initial state machine were useful.

Based on the improvements on fuzzing the FTP server, 
we answer RQ7 by concluding that there is evidence that the models learned by DICE have practical applicability to a downstream application.
DICE provides an informative state model that the fuzzer can use to guide the mutation of inputs, and 
may effectively boost the performance of a stateful server fuzzer.
In the future, we will further evaluate DICE on more protocols.

\vspace{0.2cm}\noindent\fbox{%
    \parbox{\textwidth}{%
       The FSA models learned by DICE have practical downstream application and can be used for initializing a fuzzer with a state model. 
       It improves the coverage achieved by a fuzzer on the servers of a two stateful protocols.
    }%
}

\subsection{Threats to Validity}

\textbf{Threats to internal validity.}
In our studies, we have tried to reuse most of existing implementation whenever possible. 
While it may be possible that there are bugs in the code we have written, we have checked them multiple times to reduce threats to internal validity. 
We have evaluated DICE in different aspects, such as evaluating that each component in DICE outperforms strong baselines (DSM and Tautoko for DICE, Evosuite for DICE-Tester, DSM for DICE-Miner).
We have performed a deep analysis, including a qualitative analysis, for a deeper understanding of DICE. 

\textbf{Threats to construct validity.}
In our experiments, we used common evaluation metrics that were used in previous studies. 
The evaluation process for computing these metrics has been used in previous studies and is well-understood.

\textbf{Threats to external validity.}
While it may be possible that our findings do not generalize to other library classes and APIs,  
we have considered the 11 classes evaluated in previous studies on specification mining.
These classes are diverse, coming from both the Java standard library and other third-party libraries. 
Furthermore, these are classes used in real-world software and are from a range of different domains. 
We emphasize that our approach cannot capture every possible constraint of an API or an object class, and it may not be possible to propose search goals for every constraint. 
The models DICE learned may not always be a realistic representation of every program. 
DICE is currently limited to mining finite-state automata, corresponding to regular languages. 
Still, on a benchmark created by prior studies, we have shown that the automata mined by DICE are more accurate specifications than those mined by prior approaches. 
Although DICE cannot mine specifications represented by a context-free language, 
we note many specifications mined in the literature are regular languages and researchers have found uses for these specifications, 
such as analysing and finding security flaws in bank cards~\cite{aarts2013formal} and TLS~\cite{de2015protocol}, 
or modelling Android applications~\cite{radhakrishna2018droidstar}. 
Moreover, as we have shown in Section 5.5, DICE can still learn models that are useful on a downstream task.
We leave the mining of other types of specifications, such as those equivalent to context-free languages, as future work.

\section{Related Work}
\label{sec:related_work}

We have provided an overview and explained the background of our study in Section 2, 
therefore we limit the discussion in this section. 
A key difference between our work and previous studies on specification mining is that 
we rely on an adversarial test generation process to prune incorrect properties.
For FSA inference, our work differs from existing work as we incorporate knowledge of method purity
and do not make the assumption that states with the same prefix can be merged. 

Usage models have been incorporated in test case generation previously by Fraser and Zeller~\cite{fraser2011exploiting}. 
However, they use usage models to improve the readability of test cases by guiding test generation to resemble test cases written by humans, 
and reducing the amount of nonsensical test cases generated. 
In contrast, our work aims to generate test cases that are correct, but do not resemble common usage patterns that tend to be exercised 
by existing tests. 

For the generation of test inputs,
various approaches first learn the probability distribution of observed test inputs.
While most techniques generate test inputs that resemble the learned distribution to produce synthetically correct inputs,
the Skyfire~\cite{wang2017skyfire} approach learns a probability distribution to generate test inputs that do not resemble the examples it has seen, 
applying heuristics such as favouring low probability rules.
Pavese et al.~\cite{DBLP:journals/corr/abs-1812-07525} proposed inverting the probabilities in grammar-based test generation to explore uncommon test inputs. 
Our work shares a similar goal with these two studies, aiming to explore uncommon behavior in testing.
However, our study focuses on 
a different domain of temporal specifications instead of input generation. 
Instead of using a probability distribution for modelling test inputs, we use a set of LTL property templates to characterize the execution traces while running test cases.


There has been many studies on search-based testing. 
Studies have shown that a multi-target formulation of code coverage outperforms a single-target formulation for test generation~\cite{panichella2015reformulating}. 
As discussed earlier in Section 2, Panichella et al.~\cite{panichella2017automated} proposed the DynaMOSA algorithm that enables dynamic selection of targets in 
the multi-objective problem of test case generation. 
They modelled the dependencies between structural goals, 
allowing it to outperform alternative search algorithms for search-based testing.

Many studies have studied diversity in test generation~\cite{miranda2018fast,feldt2016test,chen2010adaptive,feldt2008searching}. 
For example, researchers have studied the diversity of test inputs and outputs~\cite{alshahwan2012augmenting}.
Shin et al.~\cite{shin2016diversity} have proposed the use of diversity for mutation testing. 
Our work differs from these studies as we focus on the diversity of execution traces produced by tests, instead of 
modelling the diversity of the inputs or outputs of each method.
Moreover, we use test generation only to produce more diverse traces, 
supporting our objective of learning accurate FSA specifications.

There are other studies on the diversity of software traces, for example, coverage information collected from test execution~\cite{leon2003comparison,feldt2008searching}. 
Typically, these studies propose metrics over traces to measure the similarity of tests to maximize fault detection capability~\cite{feldt2016test}.
Our goal in this study is different, aiming to diversify traces for mining more accurate FSA specification models of a software system.

DICE is a counterexample-driven approach that has similarities to approaches based on Angluin’s L* algorithm~\cite{angluin1987learning,hungar2003domain}. 
In these studies, the oracle (aka. Minimally Adequate Teacher) provides the learner with a counterexample to answer equivalence queries. 
Using the counterexample, the learner refines the model further. 
In DICE, the DICE-Tester component can be viewed as a component providing counterexamples of LTL formulae, which is used later to construct the automata. 
There are similarities in how counterexamples to a specification (LTL formulae in DICE, the automata in L*) is used to improve the model. 
However, unlike the Minimally Adequate Teacher, the DICE-Tester is itself unaware of the ground-truth model, relying on a search-based algorithm to falsify temporal properties. 
DICE is not an active algorithm as there is no oracle interacting with the system. 
Therefore, compared to approaches based on Angluin’s L* algorithm, DICE is more practical and can work for software systems without an oracle.

There are also similarities between DICE and model checking approaches that leverage counterexamples. 
One example is CEGAR~\cite{clarke2003counterexample}, that performs counterexample-guided abstraction refinement during model checking. 
In CEGAR, counterexamples are used to split the abstract state as a counterexample indicates that there is some behaviour in the abstract model 
that is not present in the concrete version. The abstract state is split such that it no longer admits the counterexample. 
DICE is different from these approaches as it does not perform model checking, but it infers a model from a concrete system.

Adaptive model checking~\cite{groce2002adaptive} is akin to black-box model checking~\cite{peled1999black}, 
where there is no initial model of the system. 
In adaptive model checking, an inaccurate model is updated as it is used to verify a software system. 
Inaccuracies in the model may be due to differences caused by updates to the system. 
Like other Angluin-style automata learners, counterexamples are used to incrementally improve the model. 
These model checking techniques use the Vasilevskii-Chow algorithm~\cite{vasilevskii1973failure,chow1978testing}
for conformance testing to check if the model is equivalent to the system. 
DICE is similar as it tests the system against specifications, 
but differs in that it never checks or verifies conformance between the model and software system, 
which can be expensive and impractical with a cost exponential to the size of the automaton. 
DICE avoids this cost, only performing search-based testing to search for traces that falsify individual LTL properties.

\section{Conclusion and Future Work}
\label{sec:conclusion}

To conclude, we proposed a new approach of adversarial specification mining and prototyped a tool,  
DICE (\textbf{Di}versity through \textbf{C}ounter-\textbf{E}xamples), 
for mining specifications. 
DICE systematically diversifies execution traces and addresses shortcomings in current specification mining algorithms. 
By adversarially guiding test generation towards finding counterexamples of the specification, 
our approach produces diverse traces that represent uncommon but correct usage of the program.
To do so, we introduce new fitness goals representing counterexamples to temporal specifications expressed in LTL properties, 
address shortcomings in the LTL property templates used in previous studies,  
and use search-based testing to produce diverse traces. 
To take advantage of the diverse traces and the temporal properties, 
we propose a new specification mining algorithm that utilizes knowledge of method purity and 
use the temporal specifications to prevent erroneous merges to infer Finite-State Automata models 
with improved precision and recall.
Finally, in our empirical evaluation, our approach significantly outperforms 
DSM, the current state-of-the-art specification miner, and Tautoko, 
which generates tests for specification mining. 
DICE produces models with an average F-measure of 87.8, 
while the current state-of-the-art approach, DSM, produces models with an average F-measure of 68.4. 
and Tautoko produces models with an average F-measure of 53.2 in our experiments.
Furthermore, our experiments suggest that the performance of DSM does not always improve when provided with more data.
The artifact website of DICE can be found at https://kanghj.github.io/DICE

While we focus on generating uncommon sequences of method invocations in this study, we hope 
to explore the integration of methods that diversify test inputs~\cite{feldt2016test,chen2010adaptive}
to improve DICE's ability to generate uncommon test cases in future. 
We also hope to investigate more expressive LTL property types and evaluate DICE with other specifications beyond those that were studied in prior work.
We will also study the tradeoffs of including longer temporal properties in future.
Another possible direction is to explore more complex properties using temporal properties that were hard for DICE-Tester to falsify. 
The difficulty in falsifying them may indicate that the properties hold, or that there are more complex relationships between the events in the property. 
Users of DICE may also find these properties useful.
We will also study other ways to improve the effectiveness of DICE-Tester. 
To that end, we hope to explore the use of techniques such as Swarm Testing~\cite{groce2012swarm}, 
which may help to further increase the diversity of tests.

\bibliographystyle{ACM-Reference-Format}
\bibliography{testing-spec}


\begin{thebibliography}{67}


\ifx \showCODEN    \undefined \def \showCODEN     #1{\unskip}     \fi
\ifx \showDOI      \undefined \def \showDOI       #1{#1}\fi
\ifx \showISBNx    \undefined \def \showISBNx     #1{\unskip}     \fi
\ifx \showISBNxiii \undefined \def \showISBNxiii  #1{\unskip}     \fi
\ifx \showISSN     \undefined \def \showISSN      #1{\unskip}     \fi
\ifx \showLCCN     \undefined \def \showLCCN      #1{\unskip}     \fi
\ifx \shownote     \undefined \def \shownote      #1{#1}          \fi
\ifx \showarticletitle \undefined \def \showarticletitle #1{#1}   \fi
\ifx \showURL      \undefined \def \showURL       {\relax}        \fi
\providecommand\bibfield[2]{#2}
\providecommand\bibinfo[2]{#2}
\providecommand\natexlab[1]{#1}
\providecommand\showeprint[2][]{arXiv:#2}

\bibitem[\protect\citeauthoryear{Aarts, De~Ruiter, and Poll}{Aarts
  et~al\mbox{.}}{2013}]%
        {aarts2013formal}
\bibfield{author}{\bibinfo{person}{Fides Aarts}, \bibinfo{person}{Joeri
  De~Ruiter}, {and} \bibinfo{person}{Erik Poll}.}
  \bibinfo{year}{2013}\natexlab{}.
\newblock \showarticletitle{Formal models of bank cards for free}. In
  \bibinfo{booktitle}{\emph{2013 IEEE Sixth International Conference on
  Software Testing, Verification and Validation Workshops}}. IEEE,
  \bibinfo{pages}{461--468}.
\newblock


\bibitem[\protect\citeauthoryear{Alshahwan and Harman}{Alshahwan and
  Harman}{2012}]%
        {alshahwan2012augmenting}
\bibfield{author}{\bibinfo{person}{Nadia Alshahwan} {and} \bibinfo{person}{Mark
  Harman}.} \bibinfo{year}{2012}\natexlab{}.
\newblock \showarticletitle{Augmenting test suites effectiveness by increasing
  output diversity}. In \bibinfo{booktitle}{\emph{2012 34th International
  Conference on Software Engineering (ICSE)}}. IEEE,
  \bibinfo{pages}{1345--1348}.
\newblock


\bibitem[\protect\citeauthoryear{Ammons, Bod{\'\i}k, and Larus}{Ammons
  et~al\mbox{.}}{2002}]%
        {ammons2002mining}
\bibfield{author}{\bibinfo{person}{Glenn Ammons}, \bibinfo{person}{Rastislav
  Bod{\'\i}k}, {and} \bibinfo{person}{James~R Larus}.}
  \bibinfo{year}{2002}\natexlab{}.
\newblock \showarticletitle{Mining specifications}.
\newblock \bibinfo{journal}{\emph{ACM Sigplan Notices}} \bibinfo{volume}{37},
  \bibinfo{number}{1} (\bibinfo{year}{2002}), \bibinfo{pages}{4--16}.
\newblock


\bibitem[\protect\citeauthoryear{Angluin}{Angluin}{1987}]%
        {angluin1987learning}
\bibfield{author}{\bibinfo{person}{Dana Angluin}.}
  \bibinfo{year}{1987}\natexlab{}.
\newblock \showarticletitle{Learning regular sets from queries and
  counterexamples}.
\newblock \bibinfo{journal}{\emph{Information and computation}}
  \bibinfo{volume}{75}, \bibinfo{number}{2} (\bibinfo{year}{1987}),
  \bibinfo{pages}{87--106}.
\newblock


\bibitem[\protect\citeauthoryear{Arcuri and Fraser}{Arcuri and Fraser}{2013}]%
        {arcuri2013parameter}
\bibfield{author}{\bibinfo{person}{Andrea Arcuri} {and} \bibinfo{person}{Gordon
  Fraser}.} \bibinfo{year}{2013}\natexlab{}.
\newblock \showarticletitle{Parameter tuning or default values? An empirical
  investigation in search-based software engineering}.
\newblock \bibinfo{journal}{\emph{Empirical Software Engineering}}
  \bibinfo{volume}{18}, \bibinfo{number}{3} (\bibinfo{year}{2013}),
  \bibinfo{pages}{594--623}.
\newblock


\bibitem[\protect\citeauthoryear{Barnett, Naumann, Schulte, and Sun}{Barnett
  et~al\mbox{.}}{2004}]%
        {barnett200499}
\bibfield{author}{\bibinfo{person}{Mike Barnett}, \bibinfo{person}{David~A
  Naumann}, \bibinfo{person}{Wolfram Schulte}, {and} \bibinfo{person}{Qi Sun}.}
  \bibinfo{year}{2004}\natexlab{}.
\newblock \showarticletitle{99.44\% pure: Useful abstractions in
  specifications}. In \bibinfo{booktitle}{\emph{ECOOP workshop on formal
  techniques for Java-like programs (FTfJP)}}.
\newblock


\bibitem[\protect\citeauthoryear{Beschastnikh, Brun, Abrahamson, Ernst, and
  Krishnamurthy}{Beschastnikh et~al\mbox{.}}{2014}]%
        {beschastnikh2014using}
\bibfield{author}{\bibinfo{person}{Ivan Beschastnikh}, \bibinfo{person}{Yuriy
  Brun}, \bibinfo{person}{Jenny Abrahamson}, \bibinfo{person}{Michael~D Ernst},
  {and} \bibinfo{person}{Arvind Krishnamurthy}.}
  \bibinfo{year}{2014}\natexlab{}.
\newblock \showarticletitle{Using declarative specification to improve the
  understanding, extensibility, and comparison of model-inference algorithms}.
\newblock \bibinfo{journal}{\emph{IEEE Transactions on Software Engineering}}
  \bibinfo{volume}{41}, \bibinfo{number}{4} (\bibinfo{year}{2014}),
  \bibinfo{pages}{408--428}.
\newblock


\bibitem[\protect\citeauthoryear{Beschastnikh, Brun, Schneider, Sloan, and
  Ernst}{Beschastnikh et~al\mbox{.}}{2011}]%
        {beschastnikh2011leveraging}
\bibfield{author}{\bibinfo{person}{Ivan Beschastnikh}, \bibinfo{person}{Yuriy
  Brun}, \bibinfo{person}{Sigurd Schneider}, \bibinfo{person}{Michael Sloan},
  {and} \bibinfo{person}{Michael~D Ernst}.} \bibinfo{year}{2011}\natexlab{}.
\newblock \showarticletitle{Leveraging existing instrumentation to
  automatically infer invariant-constrained models}. In
  \bibinfo{booktitle}{\emph{Proceedings of the 19th ACM SIGSOFT symposium and
  the 13th European conference on Foundations of software engineering}}. ACM,
  \bibinfo{pages}{267--277}.
\newblock


\bibitem[\protect\citeauthoryear{Busany and Maoz}{Busany and Maoz}{2016}]%
        {busany2016behavioral}
\bibfield{author}{\bibinfo{person}{Nimrod Busany} {and} \bibinfo{person}{Shahar
  Maoz}.} \bibinfo{year}{2016}\natexlab{}.
\newblock \showarticletitle{Behavioral log analysis with statistical
  guarantees}. In \bibinfo{booktitle}{\emph{2016 IEEE/ACM 38th International
  Conference on Software Engineering (ICSE)}}. IEEE, \bibinfo{pages}{877--887}.
\newblock


\bibitem[\protect\citeauthoryear{Cao, Tian, Le, and Lo}{Cao
  et~al\mbox{.}}{2018}]%
        {cao2018rule}
\bibfield{author}{\bibinfo{person}{Zherui Cao}, \bibinfo{person}{Yuan Tian},
  \bibinfo{person}{Tien-Duy~B Le}, {and} \bibinfo{person}{David Lo}.}
  \bibinfo{year}{2018}\natexlab{}.
\newblock \showarticletitle{Rule-based specification mining leveraging learning
  to rank}.
\newblock \bibinfo{journal}{\emph{Automated Software Engineering}}
  \bibinfo{volume}{25}, \bibinfo{number}{3} (\bibinfo{year}{2018}),
  \bibinfo{pages}{501--530}.
\newblock


\bibitem[\protect\citeauthoryear{Chen, Kuo, Merkel, and Tse}{Chen
  et~al\mbox{.}}{2010}]%
        {chen2010adaptive}
\bibfield{author}{\bibinfo{person}{Tsong~Yueh Chen}, \bibinfo{person}{Fei-Ching
  Kuo}, \bibinfo{person}{Robert~G Merkel}, {and} \bibinfo{person}{TH Tse}.}
  \bibinfo{year}{2010}\natexlab{}.
\newblock \showarticletitle{Adaptive random testing: The art of test case
  diversity}.
\newblock \bibinfo{journal}{\emph{Journal of Systems and Software}}
  \bibinfo{volume}{83}, \bibinfo{number}{1} (\bibinfo{year}{2010}),
  \bibinfo{pages}{60--66}.
\newblock


\bibitem[\protect\citeauthoryear{Chow}{Chow}{1978}]%
        {chow1978testing}
\bibfield{author}{\bibinfo{person}{Tsun~S. Chow}.}
  \bibinfo{year}{1978}\natexlab{}.
\newblock \showarticletitle{Testing software design modeled by finite-state
  machines}.
\newblock \bibinfo{journal}{\emph{IEEE Transactions on Software Engineering}}
  \bibinfo{number}{3} (\bibinfo{year}{1978}), \bibinfo{pages}{178--187}.
\newblock


\bibitem[\protect\citeauthoryear{Clarke, Grumberg, Jha, Lu, and Veith}{Clarke
  et~al\mbox{.}}{2003}]%
        {clarke2003counterexample}
\bibfield{author}{\bibinfo{person}{Edmund Clarke}, \bibinfo{person}{Orna
  Grumberg}, \bibinfo{person}{Somesh Jha}, \bibinfo{person}{Yuan Lu}, {and}
  \bibinfo{person}{Helmut Veith}.} \bibinfo{year}{2003}\natexlab{}.
\newblock \showarticletitle{Counterexample-guided abstraction refinement for
  symbolic model checking}.
\newblock \bibinfo{journal}{\emph{Journal of the ACM (JACM)}}
  \bibinfo{volume}{50}, \bibinfo{number}{5} (\bibinfo{year}{2003}),
  \bibinfo{pages}{752--794}.
\newblock


\bibitem[\protect\citeauthoryear{Cohen and Maoz}{Cohen and Maoz}{2015}]%
        {cohen2015have}
\bibfield{author}{\bibinfo{person}{Hila Cohen} {and} \bibinfo{person}{Shahar
  Maoz}.} \bibinfo{year}{2015}\natexlab{}.
\newblock \showarticletitle{Have We Seen Enough Traces?(T)}. In
  \bibinfo{booktitle}{\emph{2015 30th IEEE/ACM International Conference on
  Automated Software Engineering (ASE)}}. IEEE, \bibinfo{pages}{93--103}.
\newblock


\bibitem[\protect\citeauthoryear{Dallmeier, Knopp, Mallon, Hack, and
  Zeller}{Dallmeier et~al\mbox{.}}{2010}]%
        {dallmeier2010generating}
\bibfield{author}{\bibinfo{person}{Valentin Dallmeier},
  \bibinfo{person}{Nikolai Knopp}, \bibinfo{person}{Christoph Mallon},
  \bibinfo{person}{Sebastian Hack}, {and} \bibinfo{person}{Andreas Zeller}.}
  \bibinfo{year}{2010}\natexlab{}.
\newblock \showarticletitle{Generating test cases for specification mining}. In
  \bibinfo{booktitle}{\emph{Proceedings of the 19th international symposium on
  Software testing and analysis}}. ACM, \bibinfo{pages}{85--96}.
\newblock


\bibitem[\protect\citeauthoryear{Dallmeier, Lindig, Wasylkowski, and
  Zeller}{Dallmeier et~al\mbox{.}}{2006}]%
        {dallmeier2006mining}
\bibfield{author}{\bibinfo{person}{Valentin Dallmeier},
  \bibinfo{person}{Christian Lindig}, \bibinfo{person}{Andrzej Wasylkowski},
  {and} \bibinfo{person}{Andreas Zeller}.} \bibinfo{year}{2006}\natexlab{}.
\newblock \showarticletitle{Mining object behavior with ADABU}. In
  \bibinfo{booktitle}{\emph{Proceedings of the 2006 international workshop on
  Dynamic systems analysis}}. ACM, \bibinfo{pages}{17--24}.
\newblock


\bibitem[\protect\citeauthoryear{de~Caso, Braberman, Garbervetsky, and
  Uchitel}{de~Caso et~al\mbox{.}}{2010}]%
        {de2010automated}
\bibfield{author}{\bibinfo{person}{Guido de Caso}, \bibinfo{person}{Victor
  Braberman}, \bibinfo{person}{Diego Garbervetsky}, {and}
  \bibinfo{person}{Sebastian Uchitel}.} \bibinfo{year}{2010}\natexlab{}.
\newblock \showarticletitle{Automated abstractions for contract validation}.
\newblock \bibinfo{journal}{\emph{IEEE Transactions on Software Engineering}}
  \bibinfo{volume}{38}, \bibinfo{number}{1} (\bibinfo{year}{2010}),
  \bibinfo{pages}{141--162}.
\newblock


\bibitem[\protect\citeauthoryear{De~Ruiter and Poll}{De~Ruiter and
  Poll}{2015}]%
        {de2015protocol}
\bibfield{author}{\bibinfo{person}{Joeri De~Ruiter} {and} \bibinfo{person}{Erik
  Poll}.} \bibinfo{year}{2015}\natexlab{}.
\newblock \showarticletitle{Protocol State Fuzzing of $\{$TLS$\}$
  Implementations}. In \bibinfo{booktitle}{\emph{24th $\{$USENIX$\}$ Security
  Symposium ($\{$USENIX$\}$ Security 15)}}. \bibinfo{pages}{193--206}.
\newblock


\bibitem[\protect\citeauthoryear{Durumeric, Li, Kasten, Amann, Beekman, Payer,
  Weaver, Adrian, Paxson, Bailey, et~al\mbox{.}}{Durumeric
  et~al\mbox{.}}{2014}]%
        {durumeric2014matter}
\bibfield{author}{\bibinfo{person}{Zakir Durumeric}, \bibinfo{person}{Frank
  Li}, \bibinfo{person}{James Kasten}, \bibinfo{person}{Johanna Amann},
  \bibinfo{person}{Jethro Beekman}, \bibinfo{person}{Mathias Payer},
  \bibinfo{person}{Nicolas Weaver}, \bibinfo{person}{David Adrian},
  \bibinfo{person}{Vern Paxson}, \bibinfo{person}{Michael Bailey},
  {et~al\mbox{.}}} \bibinfo{year}{2014}\natexlab{}.
\newblock \showarticletitle{The matter of heartbleed}. In
  \bibinfo{booktitle}{\emph{Proceedings of the 2014 Conference on Internet
  Measurement Conference}}. \bibinfo{pages}{475--488}.
\newblock


\bibitem[\protect\citeauthoryear{Dwyer, Avrunin, and Corbett}{Dwyer
  et~al\mbox{.}}{1999}]%
        {dwyer1999patterns}
\bibfield{author}{\bibinfo{person}{Matthew~B Dwyer}, \bibinfo{person}{George~S
  Avrunin}, {and} \bibinfo{person}{James~C Corbett}.}
  \bibinfo{year}{1999}\natexlab{}.
\newblock \showarticletitle{Patterns in property specifications for
  finite-state verification}. In \bibinfo{booktitle}{\emph{Proceedings of the
  1999 International Conference on Software Engineering (IEEE Cat. No.
  99CB37002)}}. IEEE, \bibinfo{pages}{411--420}.
\newblock


\bibitem[\protect\citeauthoryear{Egele, Brumley, Fratantonio, and
  Kruegel}{Egele et~al\mbox{.}}{2013}]%
        {egele2013empirical}
\bibfield{author}{\bibinfo{person}{Manuel Egele}, \bibinfo{person}{David
  Brumley}, \bibinfo{person}{Yanick Fratantonio}, {and}
  \bibinfo{person}{Christopher Kruegel}.} \bibinfo{year}{2013}\natexlab{}.
\newblock \showarticletitle{An empirical study of cryptographic misuse in
  android applications}. In \bibinfo{booktitle}{\emph{Proceedings of the 2013
  ACM SIGSAC conference on Computer \& communications security}}. ACM,
  \bibinfo{pages}{73--84}.
\newblock


\bibitem[\protect\citeauthoryear{Ernst, Perkins, Guo, McCamant, Pacheco,
  Tschantz, and Xiao}{Ernst et~al\mbox{.}}{2007}]%
        {ernst2007daikon}
\bibfield{author}{\bibinfo{person}{Michael~D Ernst}, \bibinfo{person}{Jeff~H
  Perkins}, \bibinfo{person}{Philip~J Guo}, \bibinfo{person}{Stephen McCamant},
  \bibinfo{person}{Carlos Pacheco}, \bibinfo{person}{Matthew~S Tschantz}, {and}
  \bibinfo{person}{Chen Xiao}.} \bibinfo{year}{2007}\natexlab{}.
\newblock \showarticletitle{The Daikon system for dynamic detection of likely
  invariants}.
\newblock \bibinfo{journal}{\emph{Science of computer programming}}
  \bibinfo{volume}{69}, \bibinfo{number}{1-3} (\bibinfo{year}{2007}),
  \bibinfo{pages}{35--45}.
\newblock


\bibitem[\protect\citeauthoryear{Feldt, Poulding, Clark, and Yoo}{Feldt
  et~al\mbox{.}}{2016}]%
        {feldt2016test}
\bibfield{author}{\bibinfo{person}{Robert Feldt}, \bibinfo{person}{Simon
  Poulding}, \bibinfo{person}{David Clark}, {and} \bibinfo{person}{Shin Yoo}.}
  \bibinfo{year}{2016}\natexlab{}.
\newblock \showarticletitle{Test set diameter: Quantifying the diversity of
  sets of test cases}. In \bibinfo{booktitle}{\emph{2016 IEEE International
  Conference on Software Testing, Verification and Validation (ICST)}}. IEEE,
  \bibinfo{pages}{223--233}.
\newblock


\bibitem[\protect\citeauthoryear{Feldt, Torkar, Gorschek, and Afzal}{Feldt
  et~al\mbox{.}}{2008}]%
        {feldt2008searching}
\bibfield{author}{\bibinfo{person}{Robert Feldt}, \bibinfo{person}{Richard
  Torkar}, \bibinfo{person}{Tony Gorschek}, {and} \bibinfo{person}{Wasif
  Afzal}.} \bibinfo{year}{2008}\natexlab{}.
\newblock \showarticletitle{Searching for cognitively diverse tests: Towards
  universal test diversity metrics}. In \bibinfo{booktitle}{\emph{2008 IEEE
  International Conference on Software Testing Verification and Validation
  Workshop}}. IEEE, \bibinfo{pages}{178--186}.
\newblock


\bibitem[\protect\citeauthoryear{Fraser and Arcuri}{Fraser and Arcuri}{2011}]%
        {fraser2011evosuite}
\bibfield{author}{\bibinfo{person}{Gordon Fraser} {and} \bibinfo{person}{Andrea
  Arcuri}.} \bibinfo{year}{2011}\natexlab{}.
\newblock \showarticletitle{EvoSuite: automatic test suite generation for
  object-oriented software}. In \bibinfo{booktitle}{\emph{Proceedings of the
  19th ACM SIGSOFT symposium and the 13th European conference on Foundations of
  software engineering}}. ACM, \bibinfo{pages}{416--419}.
\newblock


\bibitem[\protect\citeauthoryear{Fraser and Zeller}{Fraser and Zeller}{2011}]%
        {fraser2011exploiting}
\bibfield{author}{\bibinfo{person}{Gordon Fraser} {and}
  \bibinfo{person}{Andreas Zeller}.} \bibinfo{year}{2011}\natexlab{}.
\newblock \showarticletitle{Exploiting common object usage in test case
  generation}. In \bibinfo{booktitle}{\emph{2011 Fourth IEEE International
  Conference on Software Testing, Verification and Validation}}. IEEE,
  \bibinfo{pages}{80--89}.
\newblock


\bibitem[\protect\citeauthoryear{Gay}{Gay}{2017}]%
        {gay2017fitness}
\bibfield{author}{\bibinfo{person}{Gregory Gay}.}
  \bibinfo{year}{2017}\natexlab{}.
\newblock \showarticletitle{The fitness function for the job: Search-based
  generation of test suites that detect real faults}. In
  \bibinfo{booktitle}{\emph{2017 IEEE International Conference on Software
  Testing, Verification and Validation (ICST)}}. IEEE,
  \bibinfo{pages}{345--355}.
\newblock


\bibitem[\protect\citeauthoryear{Groce, Peled, and Yannakakis}{Groce
  et~al\mbox{.}}{2002}]%
        {groce2002adaptive}
\bibfield{author}{\bibinfo{person}{Alex Groce}, \bibinfo{person}{Doron Peled},
  {and} \bibinfo{person}{Mihalis Yannakakis}.} \bibinfo{year}{2002}\natexlab{}.
\newblock \showarticletitle{Adaptive model checking}. In
  \bibinfo{booktitle}{\emph{International Conference on Tools and Algorithms
  for the Construction and Analysis of Systems}}. Springer,
  \bibinfo{pages}{357--370}.
\newblock


\bibitem[\protect\citeauthoryear{Groce, Zhang, Eide, Chen, and Regehr}{Groce
  et~al\mbox{.}}{2012}]%
        {groce2012swarm}
\bibfield{author}{\bibinfo{person}{Alex Groce}, \bibinfo{person}{Chaoqiang
  Zhang}, \bibinfo{person}{Eric Eide}, \bibinfo{person}{Yang Chen}, {and}
  \bibinfo{person}{John Regehr}.} \bibinfo{year}{2012}\natexlab{}.
\newblock \showarticletitle{Swarm testing}. In
  \bibinfo{booktitle}{\emph{Proceedings of the 2012 International Symposium on
  Software Testing and Analysis}}. \bibinfo{pages}{78--88}.
\newblock


\bibitem[\protect\citeauthoryear{Huang and Milanova}{Huang and
  Milanova}{2012}]%
        {huang2012reiminfer}
\bibfield{author}{\bibinfo{person}{Wei Huang} {and} \bibinfo{person}{Ana
  Milanova}.} \bibinfo{year}{2012}\natexlab{}.
\newblock \showarticletitle{ReImInfer: method purity inference for Java}. In
  \bibinfo{booktitle}{\emph{Proceedings of the ACM SIGSOFT 20th International
  Symposium on the Foundations of Software Engineering}}. ACM,
  \bibinfo{pages}{38}.
\newblock


\bibitem[\protect\citeauthoryear{Hungar, Niese, and Steffen}{Hungar
  et~al\mbox{.}}{2003}]%
        {hungar2003domain}
\bibfield{author}{\bibinfo{person}{Hardi Hungar}, \bibinfo{person}{Oliver
  Niese}, {and} \bibinfo{person}{Bernhard Steffen}.}
  \bibinfo{year}{2003}\natexlab{}.
\newblock \showarticletitle{Domain-specific optimization in automata learning}.
  In \bibinfo{booktitle}{\emph{International Conference on Computer Aided
  Verification}}. Springer, \bibinfo{pages}{315--327}.
\newblock


\bibitem[\protect\citeauthoryear{Huth and Ryan}{Huth and Ryan}{2004}]%
        {huth2004logic}
\bibfield{author}{\bibinfo{person}{Michael Huth} {and} \bibinfo{person}{Mark
  Ryan}.} \bibinfo{year}{2004}\natexlab{}.
\newblock \bibinfo{booktitle}{\emph{Logic in Computer Science: Modelling and
  reasoning about systems}}.
\newblock \bibinfo{publisher}{Cambridge university press}.
\newblock


\bibitem[\protect\citeauthoryear{Krka, Brun, and Medvidovic}{Krka
  et~al\mbox{.}}{2014}]%
        {krka2014automatic}
\bibfield{author}{\bibinfo{person}{Ivo Krka}, \bibinfo{person}{Yuriy Brun},
  {and} \bibinfo{person}{Nenad Medvidovic}.} \bibinfo{year}{2014}\natexlab{}.
\newblock \showarticletitle{Automatic mining of specifications from invocation
  traces and method invariants}. In \bibinfo{booktitle}{\emph{Proceedings of
  the 22nd ACM SIGSOFT International Symposium on Foundations of Software
  Engineering}}. ACM, \bibinfo{pages}{178--189}.
\newblock


\bibitem[\protect\citeauthoryear{Le, Le, Lo, and Beschastnikh}{Le
  et~al\mbox{.}}{2015}]%
        {le2015synergizing}
\bibfield{author}{\bibinfo{person}{Tien-Duy~B Le}, \bibinfo{person}{Xuan-Bach~D
  Le}, \bibinfo{person}{David Lo}, {and} \bibinfo{person}{Ivan Beschastnikh}.}
  \bibinfo{year}{2015}\natexlab{}.
\newblock \showarticletitle{Synergizing specification miners through model
  fissions and fusions (t)}. In \bibinfo{booktitle}{\emph{2015 30th IEEE/ACM
  International Conference on Automated Software Engineering (ASE)}}. IEEE,
  \bibinfo{pages}{115--125}.
\newblock


\bibitem[\protect\citeauthoryear{Le and Lo}{Le and Lo}{2015}]%
        {le2015beyond}
\bibfield{author}{\bibinfo{person}{Tien-Duy~B Le} {and} \bibinfo{person}{David
  Lo}.} \bibinfo{year}{2015}\natexlab{}.
\newblock \showarticletitle{Beyond support and confidence: Exploring
  interestingness measures for rule-based specification mining}. In
  \bibinfo{booktitle}{\emph{2015 IEEE 22nd International Conference on Software
  Analysis, Evolution, and Reengineering (SANER)}}. IEEE,
  \bibinfo{pages}{331--340}.
\newblock


\bibitem[\protect\citeauthoryear{Le and Lo}{Le and Lo}{2018}]%
        {le2018deep}
\bibfield{author}{\bibinfo{person}{Tien-Duy~B Le} {and} \bibinfo{person}{David
  Lo}.} \bibinfo{year}{2018}\natexlab{}.
\newblock \showarticletitle{Deep specification mining}. In
  \bibinfo{booktitle}{\emph{Proceedings of the 27th ACM SIGSOFT International
  Symposium on Software Testing and Analysis}}. ACM, \bibinfo{pages}{106--117}.
\newblock


\bibitem[\protect\citeauthoryear{Legunsen, Hassan, Xu, Ro{\c{s}}u, and
  Marinov}{Legunsen et~al\mbox{.}}{2016}]%
        {legunsen2016good}
\bibfield{author}{\bibinfo{person}{Owolabi Legunsen}, \bibinfo{person}{Wajih~Ul
  Hassan}, \bibinfo{person}{Xinyue Xu}, \bibinfo{person}{Grigore Ro{\c{s}}u},
  {and} \bibinfo{person}{Darko Marinov}.} \bibinfo{year}{2016}\natexlab{}.
\newblock \showarticletitle{How good are the specs? A study of the bug-finding
  effectiveness of existing Java API specifications}. In
  \bibinfo{booktitle}{\emph{2016 31st IEEE/ACM International Conference on
  Automated Software Engineering (ASE)}}. IEEE, \bibinfo{pages}{602--613}.
\newblock


\bibitem[\protect\citeauthoryear{Lemieux, Park, and Beschastnikh}{Lemieux
  et~al\mbox{.}}{2015}]%
        {lemieux2015general}
\bibfield{author}{\bibinfo{person}{Caroline Lemieux}, \bibinfo{person}{Dennis
  Park}, {and} \bibinfo{person}{Ivan Beschastnikh}.}
  \bibinfo{year}{2015}\natexlab{}.
\newblock \showarticletitle{General LTL specification mining (t)}. In
  \bibinfo{booktitle}{\emph{2015 30th IEEE/ACM International Conference on
  Automated Software Engineering (ASE)}}. IEEE, \bibinfo{pages}{81--92}.
\newblock


\bibitem[\protect\citeauthoryear{Leon and Podgurski}{Leon and
  Podgurski}{2003}]%
        {leon2003comparison}
\bibfield{author}{\bibinfo{person}{David Leon} {and} \bibinfo{person}{Andy
  Podgurski}.} \bibinfo{year}{2003}\natexlab{}.
\newblock \showarticletitle{A comparison of coverage-based and
  distribution-based techniques for filtering and prioritizing test cases}. In
  \bibinfo{booktitle}{\emph{14th International Symposium on Software
  Reliability Engineering, 2003. ISSRE 2003.}} IEEE, \bibinfo{pages}{442--453}.
\newblock


\bibitem[\protect\citeauthoryear{Lo, Khoo, and Liu}{Lo et~al\mbox{.}}{2008}]%
        {lo2008mining}
\bibfield{author}{\bibinfo{person}{David Lo}, \bibinfo{person}{Siau-Cheng
  Khoo}, {and} \bibinfo{person}{Chao Liu}.} \bibinfo{year}{2008}\natexlab{}.
\newblock \showarticletitle{Mining temporal rules for software maintenance}.
\newblock \bibinfo{journal}{\emph{Journal of Software Maintenance and
  Evolution: Research and Practice}} \bibinfo{volume}{20}, \bibinfo{number}{4}
  (\bibinfo{year}{2008}), \bibinfo{pages}{227--247}.
\newblock


\bibitem[\protect\citeauthoryear{Lo, Mariani, and Pezz{\`e}}{Lo
  et~al\mbox{.}}{2009}]%
        {lo2009automatic}
\bibfield{author}{\bibinfo{person}{David Lo}, \bibinfo{person}{Leonardo
  Mariani}, {and} \bibinfo{person}{Mauro Pezz{\`e}}.}
  \bibinfo{year}{2009}\natexlab{}.
\newblock \showarticletitle{Automatic steering of behavioral model inference}.
  In \bibinfo{booktitle}{\emph{Proceedings of the the 7th joint meeting of the
  European software engineering conference and the ACM SIGSOFT symposium on The
  foundations of software engineering}}. ACM, \bibinfo{pages}{345--354}.
\newblock


\bibitem[\protect\citeauthoryear{Lorenzoli, Mariani, and Pezz{\`e}}{Lorenzoli
  et~al\mbox{.}}{2008}]%
        {lorenzoli2008automatic}
\bibfield{author}{\bibinfo{person}{Davide Lorenzoli}, \bibinfo{person}{Leonardo
  Mariani}, {and} \bibinfo{person}{Mauro Pezz{\`e}}.}
  \bibinfo{year}{2008}\natexlab{}.
\newblock \showarticletitle{Automatic generation of software behavioral
  models}. In \bibinfo{booktitle}{\emph{Proceedings of the 30th international
  conference on Software engineering}}. ACM, \bibinfo{pages}{501--510}.
\newblock


\bibitem[\protect\citeauthoryear{Miranda, Cruciani, Verdecchia, and
  Bertolino}{Miranda et~al\mbox{.}}{2018}]%
        {miranda2018fast}
\bibfield{author}{\bibinfo{person}{Breno Miranda}, \bibinfo{person}{Emilio
  Cruciani}, \bibinfo{person}{Roberto Verdecchia}, {and}
  \bibinfo{person}{Antonia Bertolino}.} \bibinfo{year}{2018}\natexlab{}.
\newblock \showarticletitle{Fast approaches to scalable similarity-based test
  case prioritization}. In \bibinfo{booktitle}{\emph{Proceedings of the 40th
  International Conference on Software Engineering}}. ACM,
  \bibinfo{pages}{222--232}.
\newblock


\bibitem[\protect\citeauthoryear{Molina, Kifetew, and Panichella}{Molina
  et~al\mbox{.}}{2018}]%
        {molina2018java}
\bibfield{author}{\bibinfo{person}{Urko~Rueda Molina}, \bibinfo{person}{Fitsum
  Kifetew}, {and} \bibinfo{person}{Annibale Panichella}.}
  \bibinfo{year}{2018}\natexlab{}.
\newblock \showarticletitle{Java unit testing tool competition-sixth round}. In
  \bibinfo{booktitle}{\emph{2018 IEEE/ACM 11th International Workshop on
  Search-Based Software Testing (SBST)}}. IEEE, \bibinfo{pages}{22--29}.
\newblock


\bibitem[\protect\citeauthoryear{Nadi, Kr{\"u}ger, Mezini, and Bodden}{Nadi
  et~al\mbox{.}}{2016}]%
        {nadi2016jumping}
\bibfield{author}{\bibinfo{person}{Sarah Nadi}, \bibinfo{person}{Stefan
  Kr{\"u}ger}, \bibinfo{person}{Mira Mezini}, {and} \bibinfo{person}{Eric
  Bodden}.} \bibinfo{year}{2016}\natexlab{}.
\newblock \showarticletitle{Jumping through hoops: Why do Java developers
  struggle with cryptography APIs?}. In \bibinfo{booktitle}{\emph{Proceedings
  of the 38th International Conference on Software Engineering}}. ACM,
  \bibinfo{pages}{935--946}.
\newblock


\bibitem[\protect\citeauthoryear{Naumann}{Naumann}{2007}]%
        {naumann2007observational}
\bibfield{author}{\bibinfo{person}{David~A Naumann}.}
  \bibinfo{year}{2007}\natexlab{}.
\newblock \showarticletitle{Observational purity and encapsulation}.
\newblock \bibinfo{journal}{\emph{Theoretical Computer Science}}
  \bibinfo{volume}{376}, \bibinfo{number}{3} (\bibinfo{year}{2007}),
  \bibinfo{pages}{205--224}.
\newblock


\bibitem[\protect\citeauthoryear{Pacheco and Ernst}{Pacheco and Ernst}{2007}]%
        {pacheco2007randoop}
\bibfield{author}{\bibinfo{person}{Carlos Pacheco} {and}
  \bibinfo{person}{Michael~D Ernst}.} \bibinfo{year}{2007}\natexlab{}.
\newblock \showarticletitle{Randoop: feedback-directed random testing for
  Java}. In \bibinfo{booktitle}{\emph{OOPSLA Companion}}.
  \bibinfo{pages}{815--816}.
\newblock


\bibitem[\protect\citeauthoryear{Pacheco, Lahiri, Ernst, and Ball}{Pacheco
  et~al\mbox{.}}{2007}]%
        {pacheco2007feedback}
\bibfield{author}{\bibinfo{person}{Carlos Pacheco}, \bibinfo{person}{Shuvendu~K
  Lahiri}, \bibinfo{person}{Michael~D Ernst}, {and} \bibinfo{person}{Thomas
  Ball}.} \bibinfo{year}{2007}\natexlab{}.
\newblock \showarticletitle{Feedback-directed random test generation}. In
  \bibinfo{booktitle}{\emph{Proceedings of the 29th international conference on
  Software Engineering}}. IEEE Computer Society, \bibinfo{pages}{75--84}.
\newblock


\bibitem[\protect\citeauthoryear{Panichella, Kifetew, and Tonella}{Panichella
  et~al\mbox{.}}{2015}]%
        {panichella2015reformulating}
\bibfield{author}{\bibinfo{person}{Annibale Panichella},
  \bibinfo{person}{Fitsum~Meshesha Kifetew}, {and} \bibinfo{person}{Paolo
  Tonella}.} \bibinfo{year}{2015}\natexlab{}.
\newblock \showarticletitle{Reformulating branch coverage as a many-objective
  optimization problem}. In \bibinfo{booktitle}{\emph{2015 IEEE 8th
  International Conference on Software Testing, Verification and Validation
  (ICST)}}. IEEE, \bibinfo{pages}{1--10}.
\newblock


\bibitem[\protect\citeauthoryear{Panichella, Kifetew, and Tonella}{Panichella
  et~al\mbox{.}}{2017}]%
        {panichella2017automated}
\bibfield{author}{\bibinfo{person}{Annibale Panichella},
  \bibinfo{person}{Fitsum~Meshesha Kifetew}, {and} \bibinfo{person}{Paolo
  Tonella}.} \bibinfo{year}{2017}\natexlab{}.
\newblock \showarticletitle{Automated test case generation as a many-objective
  optimisation problem with dynamic selection of the targets}.
\newblock \bibinfo{journal}{\emph{IEEE Transactions on Software Engineering}}
  \bibinfo{volume}{44}, \bibinfo{number}{2} (\bibinfo{year}{2017}),
  \bibinfo{pages}{122--158}.
\newblock


\bibitem[\protect\citeauthoryear{Pavese, Soremekun, Havrikov, Grunske, and
  Zeller}{Pavese et~al\mbox{.}}{2018}]%
        {DBLP:journals/corr/abs-1812-07525}
\bibfield{author}{\bibinfo{person}{Esteban Pavese}, \bibinfo{person}{Ezekiel~O.
  Soremekun}, \bibinfo{person}{Nikolas Havrikov}, \bibinfo{person}{Lars
  Grunske}, {and} \bibinfo{person}{Andreas Zeller}.}
  \bibinfo{year}{2018}\natexlab{}.
\newblock \showarticletitle{Inputs from Hell: Generating Uncommon Inputs from
  Common Samples}.
\newblock \bibinfo{journal}{\emph{CoRR}}  \bibinfo{volume}{abs/1812.07525}
  (\bibinfo{year}{2018}).
\newblock
\showeprint[arxiv]{1812.07525}
\urldef\tempurl%
\url{http://arxiv.org/abs/1812.07525}
\showURL{%
\tempurl}


\bibitem[\protect\citeauthoryear{Pei, Han, Mortazavi-Asl, Wang, Pinto, Chen,
  Dayal, and Hsu}{Pei et~al\mbox{.}}{2004}]%
        {pei2004mining}
\bibfield{author}{\bibinfo{person}{Jian Pei}, \bibinfo{person}{Jiawei Han},
  \bibinfo{person}{Behzad Mortazavi-Asl}, \bibinfo{person}{Jianyong Wang},
  \bibinfo{person}{Helen Pinto}, \bibinfo{person}{Qiming Chen},
  \bibinfo{person}{Umeshwar Dayal}, {and} \bibinfo{person}{Mei-Chun Hsu}.}
  \bibinfo{year}{2004}\natexlab{}.
\newblock \showarticletitle{Mining sequential patterns by pattern-growth: The
  prefixspan approach}.
\newblock \bibinfo{journal}{\emph{IEEE Transactions on Knowledge and Data
  Engineering}} \bibinfo{volume}{16}, \bibinfo{number}{11}
  (\bibinfo{year}{2004}), \bibinfo{pages}{1424--1440}.
\newblock


\bibitem[\protect\citeauthoryear{Peled, Vardi, and Yannakakis}{Peled
  et~al\mbox{.}}{1999}]%
        {peled1999black}
\bibfield{author}{\bibinfo{person}{Doron Peled}, \bibinfo{person}{Moshe~Y
  Vardi}, {and} \bibinfo{person}{Mihalis Yannakakis}.}
  \bibinfo{year}{1999}\natexlab{}.
\newblock \showarticletitle{Black box checking}.
\newblock In \bibinfo{booktitle}{\emph{Formal Methods for Protocol Engineering
  and Distributed Systems}}. \bibinfo{publisher}{Springer},
  \bibinfo{pages}{225--240}.
\newblock


\bibitem[\protect\citeauthoryear{Pham, B{\"o}hme, and Roychoudhury}{Pham
  et~al\mbox{.}}{2020}]%
        {pham2020aflnet}
\bibfield{author}{\bibinfo{person}{Van-Thuan Pham}, \bibinfo{person}{Marcel
  B{\"o}hme}, {and} \bibinfo{person}{Abhik Roychoudhury}.}
  \bibinfo{year}{2020}\natexlab{}.
\newblock \showarticletitle{AFLNET: A Greybox Fuzzer for Network Protocols}. In
  \bibinfo{booktitle}{\emph{Proceedings of the IEEE International Conference on
  Software Testing, Verification and Validation (Testing Tools Track)}}.
\newblock


\bibitem[\protect\citeauthoryear{Pnueli}{Pnueli}{1977}]%
        {pnueli1977temporal}
\bibfield{author}{\bibinfo{person}{Amir Pnueli}.}
  \bibinfo{year}{1977}\natexlab{}.
\newblock \showarticletitle{The temporal logic of programs}. In
  \bibinfo{booktitle}{\emph{18th Annual Symposium on Foundations of Computer
  Science}}. IEEE, \bibinfo{pages}{46--57}.
\newblock


\bibitem[\protect\citeauthoryear{Radhakrishna, Lewchenko, Meier, Mover,
  Sripada, Zufferey, Chang, and Cern{\`y}}{Radhakrishna et~al\mbox{.}}{2018}]%
        {radhakrishna2018droidstar}
\bibfield{author}{\bibinfo{person}{Arjun Radhakrishna},
  \bibinfo{person}{Nicholas~V Lewchenko}, \bibinfo{person}{Shawn Meier},
  \bibinfo{person}{Sergio Mover}, \bibinfo{person}{Krishna~Chaitanya Sripada},
  \bibinfo{person}{Damien Zufferey}, \bibinfo{person}{Bor-Yuh~Evan Chang},
  {and} \bibinfo{person}{Pavol Cern{\`y}}.} \bibinfo{year}{2018}\natexlab{}.
\newblock \showarticletitle{DroidStar: callback typestates for Android
  classes}. In \bibinfo{booktitle}{\emph{2018 IEEE/ACM 40th International
  Conference on Software Engineering (ICSE)}}. IEEE,
  \bibinfo{pages}{1160--1170}.
\newblock


\bibitem[\protect\citeauthoryear{Rojas, Campos, Vivanti, Fraser, and
  Arcuri}{Rojas et~al\mbox{.}}{2015}]%
        {rojas2015combining}
\bibfield{author}{\bibinfo{person}{Jos{\'e}~Miguel Rojas},
  \bibinfo{person}{Jos{\'e} Campos}, \bibinfo{person}{Mattia Vivanti},
  \bibinfo{person}{Gordon Fraser}, {and} \bibinfo{person}{Andrea Arcuri}.}
  \bibinfo{year}{2015}\natexlab{}.
\newblock \showarticletitle{Combining multiple coverage criteria in
  search-based unit test generation}. In
  \bibinfo{booktitle}{\emph{International Symposium on Search Based Software
  Engineering}}. Springer, \bibinfo{pages}{93--108}.
\newblock


\bibitem[\protect\citeauthoryear{Rojas, Fraser, and Arcuri}{Rojas
  et~al\mbox{.}}{2016}]%
        {rojas2016seeding}
\bibfield{author}{\bibinfo{person}{Jos{\'e}~Miguel Rojas},
  \bibinfo{person}{Gordon Fraser}, {and} \bibinfo{person}{Andrea Arcuri}.}
  \bibinfo{year}{2016}\natexlab{}.
\newblock \showarticletitle{Seeding strategies in search-based unit test
  generation}.
\newblock \bibinfo{journal}{\emph{Software Testing, Verification and
  Reliability}} \bibinfo{volume}{26}, \bibinfo{number}{5}
  (\bibinfo{year}{2016}), \bibinfo{pages}{366--401}.
\newblock


\bibitem[\protect\citeauthoryear{Safyallah and Sartipi}{Safyallah and
  Sartipi}{2006}]%
        {safyallah2006dynamic}
\bibfield{author}{\bibinfo{person}{Hossein Safyallah} {and}
  \bibinfo{person}{Kamran Sartipi}.} \bibinfo{year}{2006}\natexlab{}.
\newblock \showarticletitle{Dynamic analysis of software systems using
  execution pattern mining}. In \bibinfo{booktitle}{\emph{14th IEEE
  International Conference on Program Comprehension (ICPC'06)}}. IEEE,
  \bibinfo{pages}{84--88}.
\newblock


\bibitem[\protect\citeauthoryear{Shin, Yoo, and Bae}{Shin
  et~al\mbox{.}}{2016}]%
        {shin2016diversity}
\bibfield{author}{\bibinfo{person}{Donghwan Shin}, \bibinfo{person}{Shin Yoo},
  {and} \bibinfo{person}{Doo-Hwan Bae}.} \bibinfo{year}{2016}\natexlab{}.
\newblock \showarticletitle{Diversity-aware mutation adequacy criterion for
  improving fault detection capability}. In \bibinfo{booktitle}{\emph{2016 IEEE
  Ninth International Conference on Software Testing, Verification and
  Validation Workshops (ICSTW)}}. IEEE, \bibinfo{pages}{122--131}.
\newblock


\bibitem[\protect\citeauthoryear{Sun, Brown, Beschastnikh, and Stolee}{Sun
  et~al\mbox{.}}{2019}]%
        {sun2019mining}
\bibfield{author}{\bibinfo{person}{Peng Sun}, \bibinfo{person}{Chris Brown},
  \bibinfo{person}{Ivan Beschastnikh}, {and} \bibinfo{person}{Kathryn~T
  Stolee}.} \bibinfo{year}{2019}\natexlab{}.
\newblock \showarticletitle{Mining Specifications from Documentation using a
  Crowd}. In \bibinfo{booktitle}{\emph{2019 IEEE 26th International Conference
  on Software Analysis, Evolution and Reengineering (SANER)}}. IEEE,
  \bibinfo{pages}{275--286}.
\newblock


\bibitem[\protect\citeauthoryear{Thummalapenta and Xie}{Thummalapenta and
  Xie}{2009}]%
        {thummalapenta2009mining}
\bibfield{author}{\bibinfo{person}{Suresh Thummalapenta} {and}
  \bibinfo{person}{Tao Xie}.} \bibinfo{year}{2009}\natexlab{}.
\newblock \showarticletitle{Mining exception-handling rules as sequence
  association rules}. In \bibinfo{booktitle}{\emph{Proceedings of the 31st
  International Conference on Software Engineering}}. IEEE Computer Society,
  \bibinfo{pages}{496--506}.
\newblock


\bibitem[\protect\citeauthoryear{Vasilevskii}{Vasilevskii}{1973}]%
        {vasilevskii1973failure}
\bibfield{author}{\bibinfo{person}{MP Vasilevskii}.}
  \bibinfo{year}{1973}\natexlab{}.
\newblock \showarticletitle{Failure diagnosis of automata}.
\newblock \bibinfo{journal}{\emph{Cybernetics}} \bibinfo{volume}{9},
  \bibinfo{number}{4} (\bibinfo{year}{1973}), \bibinfo{pages}{653--665}.
\newblock


\bibitem[\protect\citeauthoryear{Wang, Chen, Wei, and Liu}{Wang
  et~al\mbox{.}}{2017}]%
        {wang2017skyfire}
\bibfield{author}{\bibinfo{person}{Junjie Wang}, \bibinfo{person}{Bihuan Chen},
  \bibinfo{person}{Lei Wei}, {and} \bibinfo{person}{Yang Liu}.}
  \bibinfo{year}{2017}\natexlab{}.
\newblock \showarticletitle{Skyfire: Data-driven seed generation for fuzzing}.
  In \bibinfo{booktitle}{\emph{2017 IEEE Symposium on Security and Privacy
  (SP)}}. IEEE, \bibinfo{pages}{579--594}.
\newblock


\bibitem[\protect\citeauthoryear{Xie and Notkin}{Xie and Notkin}{2003}]%
        {xie2003mutually}
\bibfield{author}{\bibinfo{person}{Tao Xie} {and} \bibinfo{person}{David
  Notkin}.} \bibinfo{year}{2003}\natexlab{}.
\newblock \showarticletitle{Mutually enhancing test generation and
  specification inference}. In \bibinfo{booktitle}{\emph{International Workshop
  on Formal Approaches to Software Testing}}. Springer,
  \bibinfo{pages}{60--69}.
\newblock


\bibitem[\protect\citeauthoryear{Yang, Evans, Bhardwaj, Bhat, and Das}{Yang
  et~al\mbox{.}}{2006}]%
        {yang2006perracotta}
\bibfield{author}{\bibinfo{person}{Jinlin Yang}, \bibinfo{person}{David Evans},
  \bibinfo{person}{Deepali Bhardwaj}, \bibinfo{person}{Thirumalesh Bhat}, {and}
  \bibinfo{person}{Manuvir Das}.} \bibinfo{year}{2006}\natexlab{}.
\newblock \showarticletitle{Perracotta: mining temporal API rules from
  imperfect traces}. In \bibinfo{booktitle}{\emph{Proceedings of the 28th
  international conference on Software engineering}}. ACM,
  \bibinfo{pages}{282--291}.
\newblock


\bibitem[\protect\citeauthoryear{Zhong and Su}{Zhong and Su}{2013}]%
        {zhong2013detecting}
\bibfield{author}{\bibinfo{person}{Hao Zhong} {and} \bibinfo{person}{Zhendong
  Su}.} \bibinfo{year}{2013}\natexlab{}.
\newblock \showarticletitle{Detecting API documentation errors}. In
  \bibinfo{booktitle}{\emph{ACM SIGPLAN Notices}}, Vol.~\bibinfo{volume}{48}.
  ACM, \bibinfo{pages}{803--816}.
\newblock


\end{thebibliography}

\end{document}